\documentclass[11pt,letterpaper,twocolumn]{article}
\usepackage[utf8]{inputenc}
\usepackage{amsmath}
\usepackage{amssymb}
 \usepackage{graphicx}
\usepackage{color}
\usepackage{url}
\usepackage{array}
\usepackage{caption} 
 \usepackage{setspace}
 \usepackage{lineno}
 
\title{\bf A Fast and Scalable   Method for \\
Inferring Phylogenetic Networks 
from \\Trees by Aligning Lineage Taxon Strings}

\author{Louxin Zhang$^{1\;*}$, Niloufar Abhari$^{2}$, Caroline Colijn$^{2}$, Yufeng Wu$^{3}$\\
\\
$^1$~Dept. of Mathematics and Centre for Data Science and Machine Learning\\
National University of Singapore, Singapore 119076\\
\\
$^2$~Dept. of Mathematics\\
Simon Fraser University, Burnaby, B.C.
Canada V5A 1S6\\
\\
$^3$~Dept. of Computer Science and Engineering\\
University of Connecticut, Storrs, CT 06269, USA\\
}
\date

\begin{document}
\onecolumn
\clearpage 
\maketitle
\thispagestyle{empty}
 * Corresponding author: Email: matzlx@nus.edu.sg; Tel: +65-65166579 
 \\
 \\
 
 {\bf Running title}: Inferring Phylogenetic Networks from Trees

\newpage
\begin{abstract}
 The reconstruction of phylogenetic networks is an important but challenging problem in  phylogenetics and  genome evolution, as the space of phylogenetic networks
is vast and cannot be sampled well. One approach to the problem is to solve 
the minimum phylogenetic network problem, in which phylogenetic trees are first inferred,  then the smallest phylogenetic network that displays all the trees is computed. The approach takes advantage of the fact that the theory of phylogenetic trees
is mature and there are excellent tools available for inferring  phylogenetic trees from a large number of
biomolecular sequences.  A tree-child network is a phylogenetic network satisfying the condition that every non-leaf node has at least one child that is of indegree one. Here, we develop a new method that infers the minimum tree-child network by 
aligning lineage taxon strings in the phylogenetic trees.  This algorithmic innovation enables us to get around the limitations of the existing programs for 
phylogenetic network inference. Our new program,  named ALTS,  is fast enough to infer a tree-child network with a large number of reticulations for a set of up to 50 phylogenetic trees with 50 taxa that have only trivial common clusters in about a quarter of an hour on average.
\end{abstract}
\setcounter{page}{1}

\section*{Introduction}

\begin{figure}[hbt!]
\centering
\includegraphics[scale=0.6]{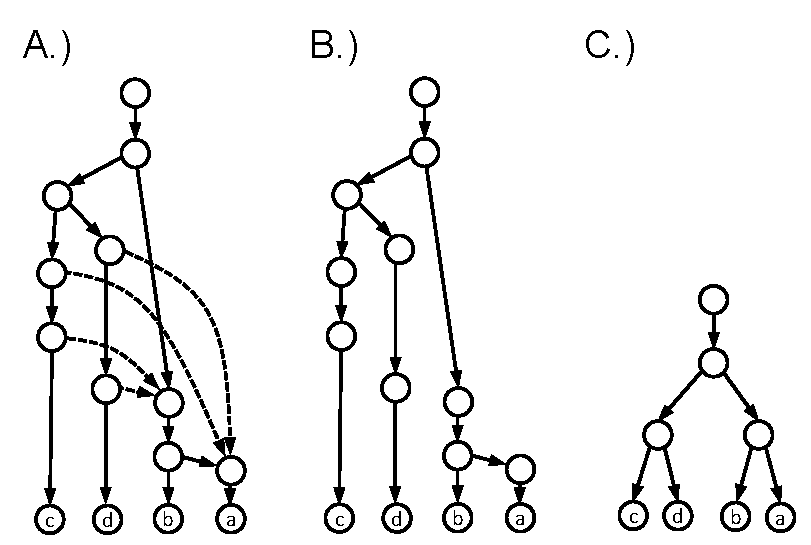}
\caption{\small 
(A.)  A tree-child network with two reticulate nodes on the taxa (a to d).  {(B.)} A subtree that was obtained by the removal of  the dashed incoming edges  of the reticulate nodes in the network. {(C.)} A tree displayed in the network, which was obtained from the subtree in (B) by removing all degree-2 nodes through combining their unique incoming and outgoing edges into an edge.
\label{display} }
\end{figure}

In this study, phylogenetic networks over a set of taxa are rooted,  directed acyclic graphs in which leaves represent the taxa,  the non-leaf indegree-1 nodes  represent speciation events  and the nodes with multiple incoming edges represent reticulation events.
The non-leaf indegree-1 nodes are called tree nodes;  the other non-leaf nodes are called reticulate nodes.
 We assume that each tree node is of outdegree 2; each reticulate node and the network root is of outdegree 1 in a phylogenetic network (Figure~\ref{display}). 
 Phylogenetic trees are  just phylogenetic networks with no reticulate nodes and thus are binary. 
 (Basic concepts and notation can be found in 
 the Supplemental Methods.) 

Now that  a variety of genomic projects have been completed, reticulate evolutionary events (e.g. horizontal gene transfer, introgression and hybridization) have been demonstrated to play important roles in genome evolution (Fontaine et al. 2015; Gogarten and Townsend 2005; Koonin et al. 2001; Marcussen et al. 2014). 
Although  phylogenetic networks are appealing for modeling  reticulate events (Koblm{\"u}ller et al. 2007), it is extremely challenging to apply phylogenetic networks in the study of genome evolution. One reason for this is that a computer program has yet to be made available for analyzing data as large as what current research is interested in (Molloy et al. 2021; Wu 2020), although recently, Bayesian methods have been used to reconstruct reassortment networks, which describe patterns of ancestry in which lineages may have different parts of their genomes inherited from distinct parents (M{\"u}ller et al. 2020; M{\"u}ller et al. 2022).  

Here, we focus on 
computing  phylogenetic networks that display a  given set of gene trees (Albrecht et al. 2012; Elworth et al. 2019;  van Iersel et al. 2022; Whidden et al. 2013; Wu 2010).   
In this approach,  trees are first inferred from biomolecular sequences and then used to reconstruct a phylogenetic network with the smallest hybridization number (HN) that displays all the trees  (see Elworth et al. 2019), where the HN  is defined as the sum over all the reticulate nodes  of the indegree of each reticulate node minus 1. This approach takes  advantage of the fact that the theory of phylogenetic trees is mature and there are excellent tools available for inferring trees from  a large number of sequences.  It has been used in evolutionary studies
(Koblm{\"u}ller et al. 2007; Marcussen et al. 2014).

Although this parsimonious approach is faster than the maximum likelihood approach (Lutteropp et al. 2022),
the  parsimonious network inference problem is still NP-hard  even for the special case when there are only two input trees (Bordewich and Semple 2007).  For the two-tree case, the fastest programs include MCTS-CHN (Yamada et al. 2020) and HYBRIDIZATION NUMBER (Whidden et al. 2013). For the general case where there are multiple input trees,  HYBROSCALE (Albrecht 2015) and its predecessor (Albrecht et al. 2012), 
PRIN (Wu 2010) and PRINs (Mirzaei and Wu 2015),  have been developed.  
All of these methods reconstruct a tree-child network with the smallest HN. Some of the methods insert reticulate edges or use other editing operations to search a network in the network space. Others reduce the tree-child network reconstruction problem to finding maximum acyclic agreement forests for the  set input trees. Finally, some methods combine both of these techniques. 
Unfortunately, none of them  will work for inferring a network from  more than 30 trees if the trees have 30 or more taxa and do not have any non-trivial taxon clusters in common,  where a non-trivial taxon cluster of a tree consists of all taxa below a tree node that is neither a leaf nor the root.

 Since the network space is vast and cannot be fully sampled, attention has been switched to the inference of tree-child networks (Cardona et al. 2009),  in which every non-leaf node has at least one child that is  not reticulate,  or, recently,  a tree-based network (Pickrell and Pritchard 2012). Tree-child network is a superclass of phylogenetic trees with a completeness property that for any set of phylogenetic trees,  there exists always a tree-child network  that displays all the trees (Linz and Semple 2019). Other desired properties of tree-child networks include the fact that
 all the tree-child networks are efficiently enumerated (Zhang 2019; Cardona and Zhang 2020). 

%
 
\section*{Results}

We mainly report a scalable computer program for inferring tree-child networks from multiple gene trees. The our program  ALTS takes a different approach that reduces the network inference problem to  aligning the lineage taxon strings computed from the input trees with respect to (w.r.t.) an ordering on the taxa. 
 
\subsection*{The inference algorithm}

Consider a set $X$ of taxa.
Let $T$ be a binary phylogenetic tree on  $X$ and let $N$ be a  tree-child network on $X$. $N$ displays
$T$ if $T$ can be obtained from $N$ by (i) removing all but one incoming edge for each reticulation node and then (ii) deleting  all degree-2 nodes (which were reticulation nodes in $N$)  (Figure~\ref{display}).

 The inference algorithm we introduce here will check all possible orderings on the taxon set to obtain the tree-child networks with the smallest HN (and equivalently the smallest number of nodes).
Let $X$ be a taxon set such that $|X|=n$ and let  $\pi=\pi_1\pi_2\cdots \pi_n$, representing a (total) ordering of $X$,  by which $\pi_i$ is `less than'  $\pi_{i+1}$ for each $i< n$. For any non-empty subset $X'$ of $X$, we use $\min_\pi(X')$ and $\max_\pi(X')$ to denote the minimum and maximum taxon of $X'$  
w.r.t. $\pi$, respectively. 

\begin{figure}[b!]
\centering
\includegraphics[scale=0.5]{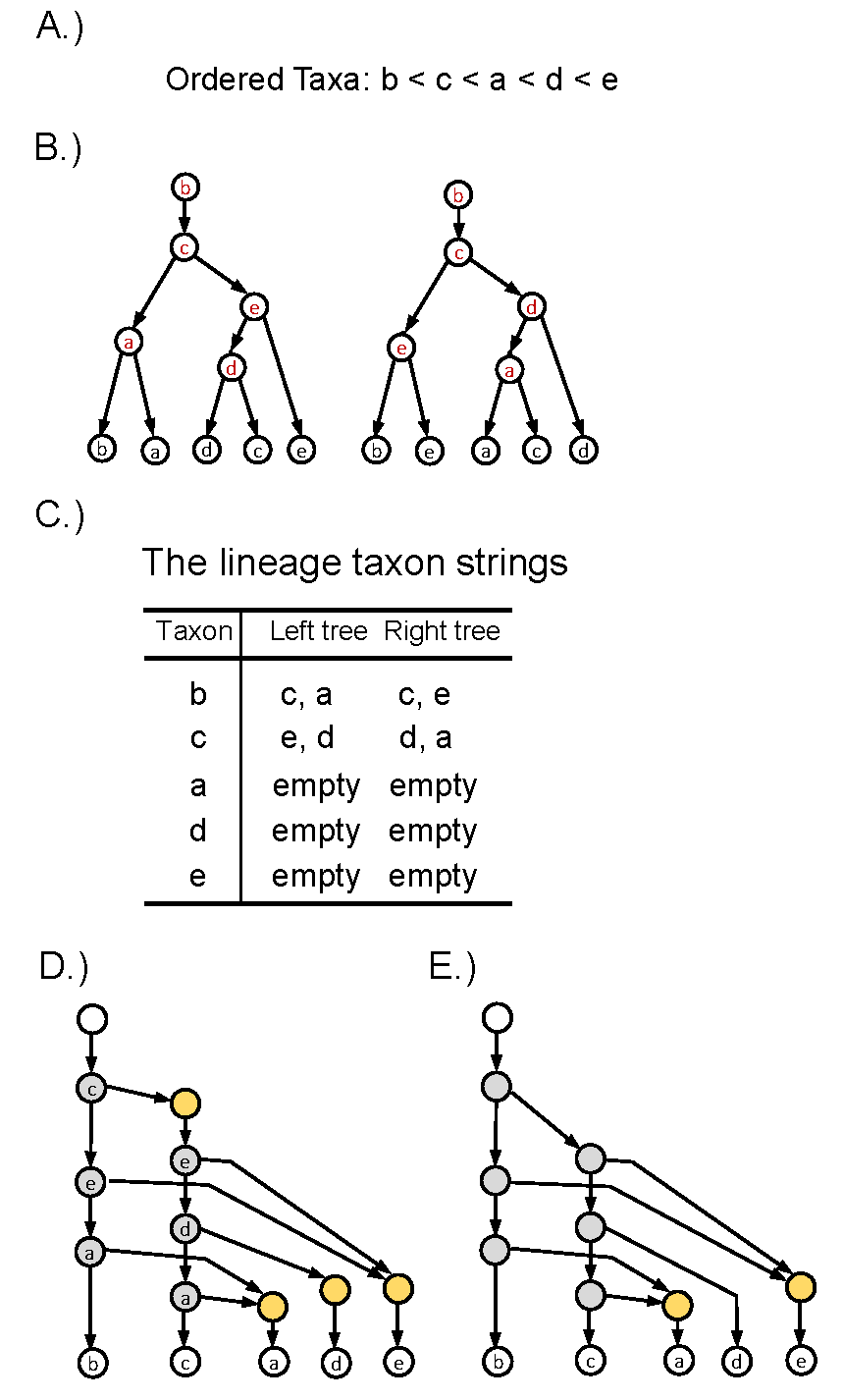}
\caption{\small The construction of a tree-child network that displays two phylogenetic trees. 
(A) An ordering on $\{a, b, c, d, e\}$. (B) Two trees, where the internal nodes are labeled w.r.t. the ordering using the {\sc Labeling} algorithm. (C) The lineage taxon strings (LTSs) of the taxa obtained from the labeling in Panel B. (D) The rooted directed graph constructed from the shortest common supersequences (SCS) of the LTSs of the taxa (in Panel C) using  {\sc Tree-child Network Reconstruction}. The SCS is $[c, e, a]$ for $[c, a]$ and $[c, e]$,  and is $[e, d, a]$ for  $[e, d]$ and $[d, a]$.
(E) The tree-child network obtained after the removal of the degree-2 nodes. \label{Fig2_infer}} 
\end{figure}

Since the root of $T$ is of outdegree 1,  $T$ has $n$ non-leaf nodes, called internal nodes.  We label the  $n$ internal  nodes of $T$  one-to-one with the taxa w.r.t. $\pi$  
by assigning the smallest taxon to the degree-1 root and assigning $\max_{\pi}\{t_v, t_w\}$ to an internal node with children $v$ and $w$, where $t_v$ is the smallest taxon below $v$ ({\sc Labelling}, Supplemental Methods).
%
For instance, let $X=\{a, b, c, d, e\}$ and $\pi=bcade$ (Figure~\ref{Fig2_infer}A). The two trees on $X$ in Figure~\ref{Fig2_infer}B have their internal nodes  labeled w.r.t. $\pi$  using {\sc Labeling}. 

Let $\tau$ be a specific taxon of $X$ such that $\tau\neq \pi_1$. We consider the unique path from the root $\rho$ to the leaf $\ell$ that represents $\tau$ in $T$:
$u_0=\rho, u_1, u_2, \cdots, u_k=
\ell.$ 
Then, $\min_{\pi}(C(u_k))=\tau$, whereas 
$\min_{\pi}(C(u_0))=\min_{\pi}(C(u_1))=\pi_1 <_{\pi} \tau$. 
Since $\min_{\pi}(C(u_k))\leq_{\pi} \min_{\pi}(C(u_{k+1}))$, there is a unique index $j$ such that $1\leq j<k$ and $\min_{\pi}(C(u_j))<\tau= \min_{\pi}(C(u_{j+1}))$. 
This implies that $u_j$ was labeled with $\tau$ by applying {\sc Labelling} and no other internal node got the same label.
The  sequence  consisting of the labels of $u_{j+1}, u_{j+2}, \cdots, u_{k-1}$ is called 
{\it the lineage taxon string} (LTS) of  $\tau$.
 The 
LTSs computed in the trees given in Figure~\ref{Fig2_infer}B  are listed in Figure~\ref{Fig2_infer}C. 

Conversely, for the LTS  of each taxon $\tau$, we construct a directed path  whose nodes are labeled one-to-one with the taxa of the LTS and add a leaf labeled with $\tau$ below the path. After we connect the first node of the resulting path ending with each taxon other than $\pi_1$ to all the nodes labelled with the taxon in other paths, we obtain $T$.
Thus, the LTSs obtained from  $T$ under any ordering $\pi$  on $X$ can be used to recover uniquely $T$. 

 A string $s$ is said to be a common supersequence of multiple strings if  all the strings can be obtained from $s$ by erasing zero or more symbols. 
Let  $\{T_1, T_2, \cdots, T_k\}$  be a set of $k$ trees on $X$.
 Let $\alpha_{ji}$ be the 
 LTS of $\pi_i$ in $T_j$ for each $i$ from 1 to $n$. (Note that 
 $\alpha_{jn}$ is the empty string for each $j$.)  Assume that, for each $i$, $\beta_i$ is a common supersequence of all $\alpha_{1i}, \alpha_{2i}, \cdots, \alpha_{ki}$ on $X$.  We can construct a tree-child network $N_\pi(\beta_1,\beta_2, \cdots, \beta_{n-1})$ on $X$ using the {\sc Tree-Child Network Construction} algorithm given below.
 \begin{center}
 \begin{tabular}{l}
 \hline
 {\sc Tree-Child Network Construction} \\
  1. ({\bf Vertical edges}) For each $\beta_i$,  define a 
  path $P_i$ with $\vert \beta_i\vert +2$ nodes: \\
  \hspace*{3em} $h_i, v_{i1}, v_{i2}, \cdots, v_{i\vert\beta_i\vert}, \ell_{\pi_i}$,\\
  \hspace*{2em}where $\beta_n$ is the empty sequence.\\
  2.  ({\bf Left--right edges}) 
  Arrange the $n$ paths
  from left to right as $P_1, P_2, \cdots, P_n$. \\ 
  \hspace*{2em}If the $m$-th symbol of $\beta_i$ is $\pi_j$, we add an
  edge $(v_{im}, h_{j})$ for each $i$ and each $m$. \\
  3. For each $i>1$, if $h_i$ is of indegree 1, eliminate $h_i$  by removing $h_i$,  together with   \\
  \hspace*{2em}its incoming and outgoing edge,  and adding a new edge from its parent to its child. \\
    \hline
 \end{tabular}
 \end{center}

 The algorithm is illustrated in Figure~\ref{Fig2_infer}D, where the SCSs are
 $[c,e,a]$ and $[e,d,a]$ for 
 $\pi_1=b$ and $\pi_2=c$,  and the empty sequence for $\pi_3=a$ and $\pi_4=d$.

The  network output from {\sc Tree-Child Network Construction} is always a  tree-child network (Proposition 2, Supplemental Methods).
Combining {\sc Labelling} and {\sc Tree-Child Network Construction}, we obtain the following exact algorithm for the network inference problem, for which the correctness is proved in Section A of the Supplemental Methods. 

\begin{center}
\begin{tabular}{l}
\hline 
  {\sc Algorithm A}     \\
  {\bf Input}:  $K$ trees $T_1, T_2, \cdots, T_k$ on $X$,  $|X|=n$.\\
  0. Set $M=\infty$ and define $n-1$ string variables $S_1, S_2, \cdots, S_{n-1}$;\\
  1.  For each ordering $\pi=\pi_1\pi_2 \cdots \pi_n$ on $X$: \\
  \hspace*{1em} 1.1. Call {\sc Labeling} to 
  label the internal nodes in each $T_i$;\\
  \hspace*{1em} 1.2.  For each taxon $\pi_j$, compute its 
   LTS $s_{ij}$ in each  $T_i$;\\
  \hspace*{1em} 1.3.  Compute the SCS $s_j$ of $s_{1j}, s_{2j}, \cdots, s_{kj}$  for each $j<n$;\\
  \hspace*{1em} 1.4. If $M> \sum_{j=1}^{n-1} |s_j|$, update $M$ to  the length sum; update $S_j$ to $s_j$ for each $j$;\\
  2. Call 
   {\sc Tree-Child Network Construction} to compute 
   a tree-child network\\
   \hspace*{1em} from the strings $S_1, S_2, \cdots, S_{n-1}$.\\
  \hline
\end{tabular}
\end{center}

\subsection*{A scalable version} 

Since there are $n!$ possible orderings on $n$ taxa and 15! is already too large, {\sc Algorithm A} is not fast enough for a set of multiple trees on 15 or more taxa.   
Another obstacle to scalability  is computing the SCS for the LTS
of each taxon. 
We achieved high scalability by using an ordering sampling method and a progressive approach for the SCS problem. 

First, the ordering sampling starts with an arbitrary ordering on the taxa and finishes in $\lfloor n/2\rfloor$ iterative steps. Assume that $\Pi_m$ is the set of orderings obtained in the $m$-th step ($m\geq 1$) such that  $\vert \Pi_m \vert \leq H$ for a parameter $H$ predefined to bound the running time.  In the $(m+1)$ step, 
for each ordering $\pi=\pi_1\pi_2\cdots \pi_n \in \Pi_m$, we generate $(n-2m+1)(n-2m)$ new orderings by interchanging
$\pi_{2m-1}$ with $\pi_{i}$ and interchanging $\pi_{2m}$ with $\pi_j$ for every possible $i$ and $j$ such that $i\neq j$,
$i> 2m$ and $j> 2m$. For each new ordering $\pi'=\pi'_1\pi_2'\cdots \pi'_n$, 
we compute a SCS $s_i$ of the 
LTSs of Taxon $\pi'_i$ in the input trees for each $i\leq 2m$. 
We compute $\Pi_{m+1}$ by sampling at most $H$ new orderings that have the smallest length sum $\sum_{1\leq i\leq 2m} |s_i|$.

Second, different progressive approaches can be used to compute a short common supersequence for 
LTSs in each sampling step (Fraser 1995). We use the following approach:
\begin{quote}
    A common supersequence  of $n$ strings  is computed in $n-1$ iterative steps. In each step, a pair of strings $s_i$ and $ s_j$ such that the SCS of $s_i$ and $s_j$,  $\mbox{SCS}(s_i, s_j)$,  has the minimum length,  over all possible string pairs,  is selected and replaced with
    $\mbox{SCS}(s_i, s_j)$.  
\end{quote}

Although the above algorithm had good performance for our purpose according to our test, it cannot always output the shortest
solution for all possible instances. The reason is that finding the SCS for arbitrary strings is NP-hard in general (Gary and Johnson 1979) and our algorithm is as a
linear-time algorithm unlikely to be the exact algorithm.

After the sampling process finishes, we obtain a set $\Pi_{\lfloor n/2\rfloor}$ of good ordering; for each ordering, we obtain 
a short common supersequence of the 
LTSs of a taxon obtained from the input trees. 
To further improve the tree-child network solution, we also use the dynamic programming algorithm to recalculate a short common supersequence  for the LTSs of each taxon,   subject to the 1G memory usage limit. We then use whichever is shorter to compute a tree-child network. 

\subsection*{Implementation of the algorithm}
\label{sect3.3}

Another technique for improving the scalability is
to decompose the input tree set into irreducible sets of trees if the input trees are reducible (Albrecht et al. 2012; Wu 2010) (Section C,  Supplemental Methods). Here, a set of trees are reducible if there is at least one common cluster except the singletons and the whole taxon set.

 Our program is named ALTS,  an acronym for ``Aligning Lineage Taxon Strings".  It can be downloaded from the GitHub site (see Software Availability).  We also developed a program that assigns a weight to each edge of the obtained tree-child network if the input trees are weighted (Section C,  Supplemental Methods). 
 
 In summary, the process of reconstructing a parsimonious  tree-child network involves the following steps.
 (i) Decompose the input tree set $S$ into irreducible tree sets, say
   $S_1, S_2, \cdots, S_t$.
  (ii)  Infer a set $N_i$ of tree-child networks for each $S_i$.
   (iii) Assemble the tree-child networks in $N_1, N_2, \cdots, N_t$ to obtain the networks that display all the trees in $S$.
   (iv) If the input trees are weighted, the branch weights are estimated for the output tree-child networks.

\subsection*{Validation experiments}
\label{sec_validation}

We assessed the accuracy and scalability of ALTS on a collection of simulated datasets that were generated using an approach reported in Wu (2010)  (See Methods section). 
\\

\noindent {\bf The optimality evaluation}~~
We compared ALTS with two heuristic network inference programs: PRINs (Mirzaei and Wu 2015), which infers an arbitrary phylogenetic network, and  van Iersel et al.'s method (van Iersel et al. 2022), which infers a tree-child network. 
We first ran the three methods  on 
50 sets of trees on 20 and 30 taxa,  each containing 10 trees.
Van Iersel et al.'s program is a parallel program. It could run successfully only on 44 (out of 50) tree sets in the 20-taxon case and 27 (out of 50) tree sets in the  30-taxon case. It was aborted for the remaining datasets after  24 hours of clock time (or about 1000 CPU hours) had elapsed. 

ALTS output tree-child networks with the same HN as  van Iersel et al.'s method on all but three datasets where the latter ran successfully. 
  The HN of the tree-child networks inferred with ALTS was one more than
  that inferred with the latter on two 20-taxon 10-tree datasets and three more than the latter on one 30-taxon 10-tree dataset. 
  Moreover, Van Iersel et al.'s method only outputted  a tree-child network, whereas ALTS computed multiple tree-child networks with the same HN.  

PRINs ran successfully on 49 out of 50 datasets in the 20-taxon case.  
In theory, the HN is inherently equal to or less than the HN of the optimal tree-child networks for every tree set. In the 20-taxon 10-tree case, the tree-child HN  inferred with ALTS was 
equal to that inferred with PRINs on 20 datasets.
The 29 discrepancy cases are summarised in the first row of Table~\ref{Optimality_summary_1}.
In the 30-taxon case, the HN difference of the two programs was also at most four (Row 2, Table~\ref{Optimality_summary_1}). The tree-child HN inferred by ALTS was even one less than the HN inferred by PRINs on one dataset.
%

\begin{table}[t!]
\caption{Summary of the HN discrepancy between 
ALTS and PRINs in 20-taxon and 30-taxon datasets each containing 10 trees. \label{Optimality_summary_1}}
\renewcommand\arraystretch{1.3}
\begin{tabular}{|l|*{5}{>{\centering}p{5mm}}p{5mm}|}
\hline
    & \multicolumn{6}{c|}{ $\mbox{HN}_{\rm ALTS}$ minus $\mbox{HN}_{\rm PRINs}$}                                            \\
{\bf Date type} & -1& 0 & 1 & 2 & 3 & 4\\
\hline
20-taxon trees  &  & 20 & 11   &   9  &   6  &   3   \\
30-taxon trees  &  1 & 5& 13  &   14  &   16  &   1  \\
\hline  
\end{tabular}
\end{table}


In summary, ALTS is almost as accurate as  van Iersel et al.'s method in terms of minimizing network HN.
The comparison between ALTS and PRINs indicated that the tree-child HN is rather close to the HN for multiple trees when the number of taxa is not too big.
\\

\begin{figure}[t!]
\centering
\includegraphics[scale=0.6]{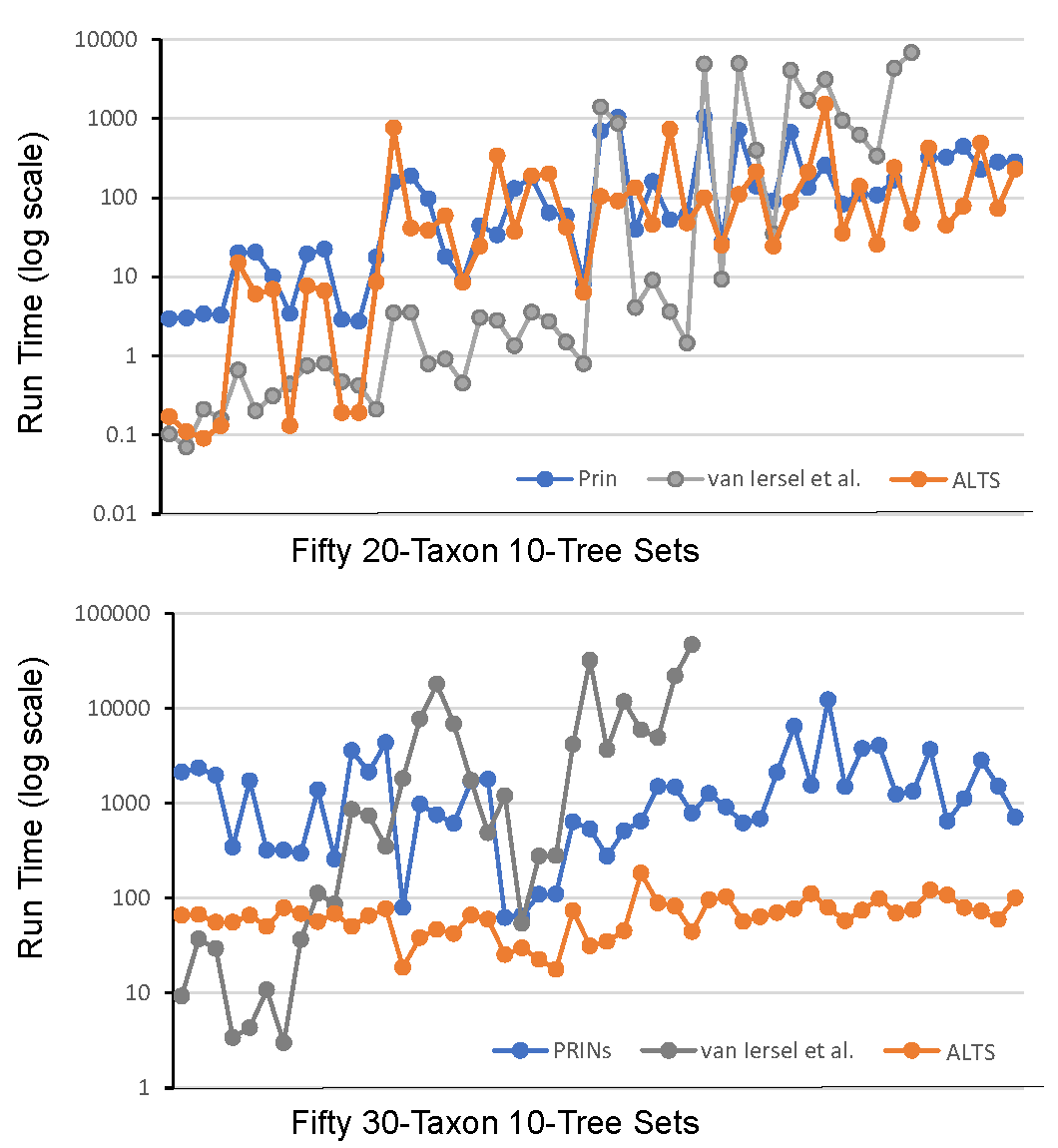}
\caption{Run time (in seconds) of the three methods on 100 datasets, each containing 10 trees on 20 or 30 taxa. The datasets are sorted in the increasing order according to the HN output from  PRINs.  Van Iersel et al.'s method had some missing data points due to an abort during the running time.  \label{Fig7_time}}
\end{figure}

\begin{figure}[t!]
\centering
\includegraphics[scale=0.55]{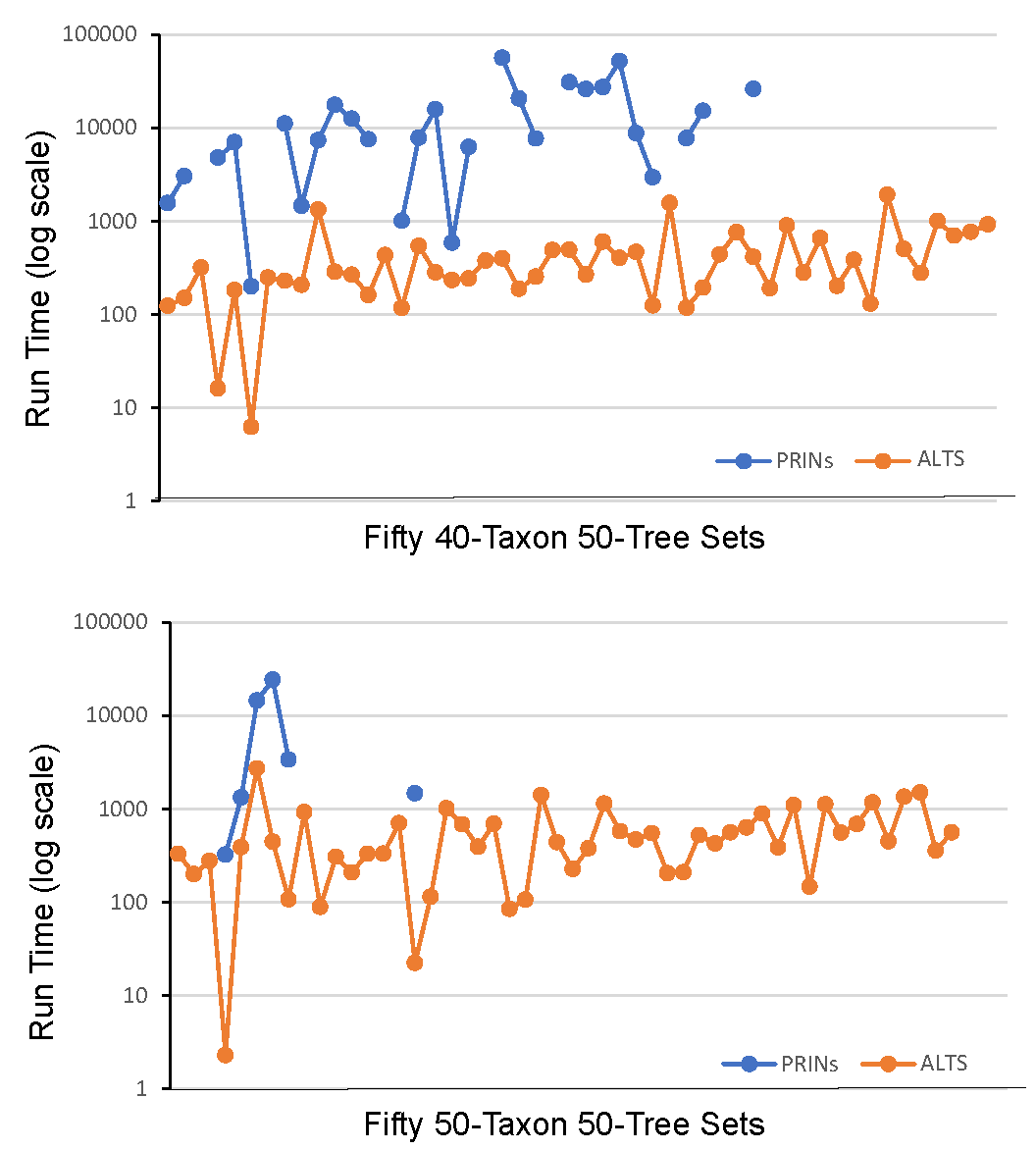}
\caption{The run time (in seconds) of ALTS on 100 datasets, each containing 50 trees on 40 or 50 taxa. The datasets are sorted in the increasing order according to the HN of the tree-child networks inferred by ALTS. \label{Fig8_time}}
\end{figure}

\noindent {\bf The scalability evaluation}~ The wall-clock time of the three methods on 100 datasets, each having 10 trees on 20 or 30 taxa,  are summarized in Figure~\ref{Fig7_time}. 
In the 20-taxa 10-tree case, the HN inferred by PRINs ranged from 5 to 17.   ALTS finished in 0.09 s to 25 m 14 s (with the mean being 2 m 21 s).  On the 49 (out of 50) 20-taxa 10-tree datasets on which PRINs finished, it took 2.94 s to 17 m 19 s (with the mean being 2 m 58 s).  ALTS was faster than PRINs on 35 tree sets. On average,  PRINs and ALTS were comparable in time for this case.

On the 44 20-taxa 10-tree datasets on which  van Iersel et al.'s method finished, its run time ranged from 
0.07 s to 82 m 22 s (with the mean being 13 m 3 s). Van Iersel et al.'s method ran faster than ALTS on 26 datasets where the HN inferred by PRINs was less than 11. One reason for this is probably that the former is a parallel program.
However, ALTS was faster than  van Iersel et al.'s method on the remaining  18 tree sets where the HN inferred by PRINs was 12 or more.

In the 30-taxon case, the HN of the solution from PRINs ranged from 8 to 21.
As shown in Figure~\ref{Fig7_time}, ALTS was faster than 
PRINs on every dataset. 
Van Iersel et al.'s method  finished on 31 (out of 50) datasets, for which  the HN of the solution obtained with PRINs was 15 or more.
ALTS was faster than Van Iersel et al.'s method  on 23 datasets, whereas
Van Iersel et al.'s method  was faster than ALTS on the remaining 8 datasets.
On average, in the 30-taxon case
 ALTS was 24 and 53 times faster than PRINs and the van Iersel et al.'s method, respectively.

Lastly,  we further ran ALTS on 100 datasets,  each containing 50 trees on  40  or  50 taxa. PRINs finished on twenty-eight  40-taxon 50-tree datasets and  five 50-taxon 50-tree datasets.  In the 40-taxon 50-tree case, 
ALTS finished in 3 s to 31 m 52 s (with the mean being 7 m 14 s). 
On contrast, 
PRINs finished on 28 tree sets, taking 3 m 19 s to 15 h 34 m 52 s (with the mean being 3 h 49 m 46 s)
(Figure~\ref{Fig8_time}).

In the 50-taxon 50-tree case,   ALTS finished in 2 s to 45 m 12 s (with the mean being 9 m 24 s) (Figure~\ref{Fig8_time}). In contrast,  van Lersel et al's method could not finish on any irreducible set of 50 trees on 50 taxa. 
PRINs finished on five tree sets in 2 h 25 m on average (Figure~\ref{Fig8_time}).

Taken together, these results suggest that ALTS has high scalability and is fast enough to infer tree-child networks  for large  tree sets.
\\

\begin{figure}[t!]
\centering
\includegraphics[scale=0.6]{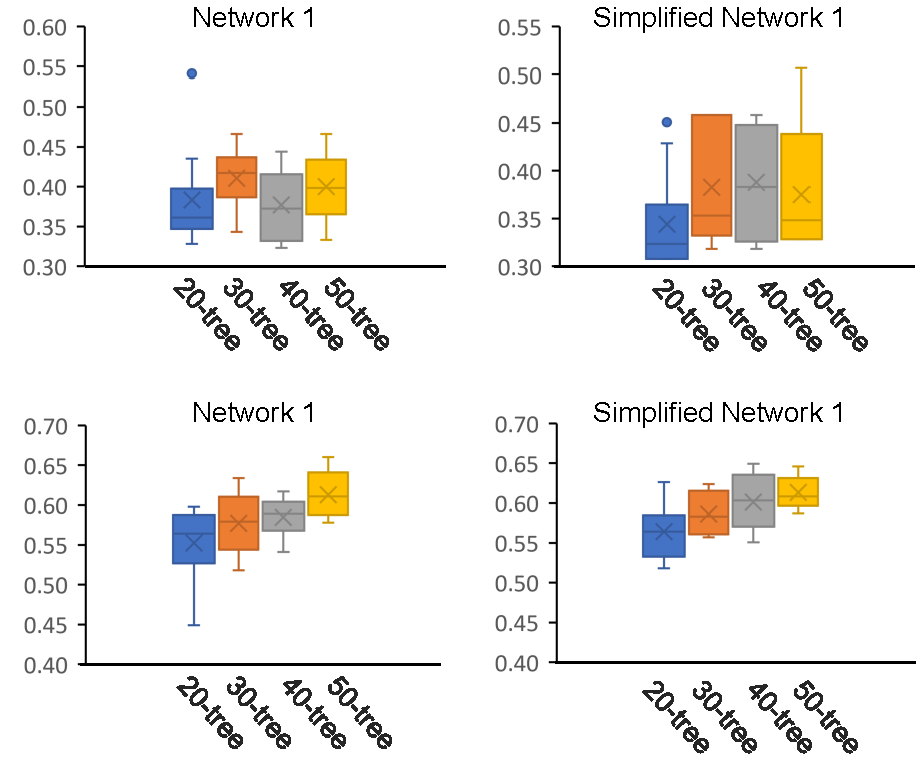}
\caption{The  box and whisker plots for the dissimilarity scores for the original networks (Supplemental Fig. S1 and S2) and one inferred by ALTS in four cases. In each plot, the four bars from left to right summarize the dissimilarity scores for the original network and 10 networks inferred from 20-, 30-, 40-, and 50-tree sets, respectively. 
\label{fig_dissimilarity}}
\end{figure}

\noindent {\bf The accuracy evaluation}~~
Evaluating the accuracy of  ALTS (and the other two methods) is not straightforward. The random networks that were used to generate the tree sets used in the last two subsections 
are not tree-child networks and contain frequently a large number of deep reticulation events. On the other hand, by the principle of parsimony, the networks inferred by the three programs contain far less number of reticulation events.  As such, 
we assessed the accuracy of ALTS by using a Jaccard distance that measures the symmetric difference between the set of clusters in the original networks and in the network inferred by ALTS (Huson et al. 2010) (see Methods).

We considered two simulated networks containing 16 binary reticulations (Network 1, Supplemental Fig.~S1) and 19 binary reticulations (Network 2,  Supplemental Fig.~S2). 
The two networks were produced using the same simulation program as used for the optimality evaluation but with a lower ratio of reticulation events. 
We also examined a simplified version of the two networks that were obtained by merging a reticulate node and its child if the reticulate node has a unique child and the child is also a reticulation node.   The two simplified networks have 9 and 10  reticulation events, respectively (bottom, Supplemental Fig.~S1 and S2).  For each network and each $k=20,30, 40, 50$, we generated 10 $k$-tree sets. 
For each tree set, we inferred a network using ALTS and computed the dissimilarity score for it and the original network. The dissimilarity score analyses are summarised in Figure~\ref{fig_dissimilarity}.

Network 1 (and its simplified version) contains less reticulation events than Network 2. We had slight better reconstruction accuracy for Network 1 than Network 2 (mean dissimilarity score range [0.3, 0.45] vs. [0.55, 0.65], Figure~\ref{fig_dissimilarity}). Also, the reconstruction from the trees sampled from each network was not significantly better than that from its simplified version. Given that all four networks can contain as many as $2^{17}$ trees,  the  results suggest that 50 trees are far less than enough for accurate reconstruction of both non-binary networks. 

On the other hand, ALTS performed well for inferring a binary tree-child network with 13 binary reticulation nodes on 22 taxa. We sampled trees from the binary tree-child network given in Supplemental Fig.~S3. We could reconstruct  the network on  1 out of 10 random 5-tree sets, 6 out of 10 random 10-tree sets and all 10 random 20-tree sets.

\begin{figure}[bt!]
\centering
\includegraphics[scale=0.5]{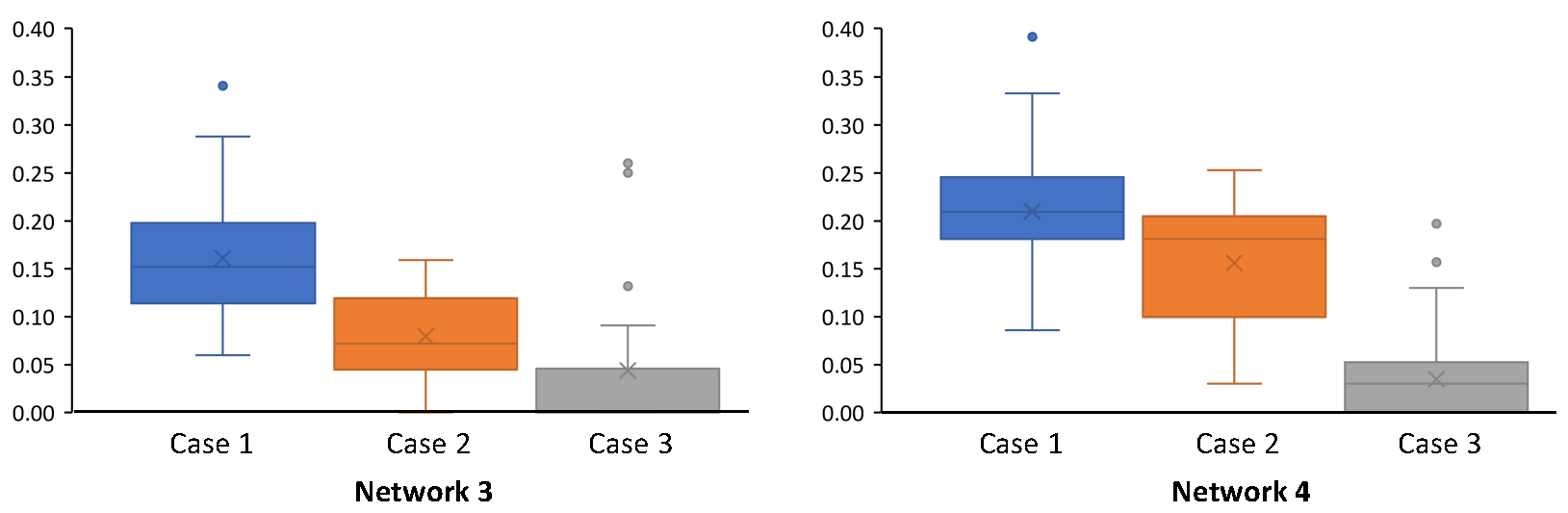}
\caption{ The  box and whisker plots for the dissimilarity scores for the original networks (Supplemental Fig.~S4) and the network inferred using ALTS in three cases. In each plot, the three bars from left to right summarize the 50 dissimilarity scores obtained using 5 random inferred trees (Case 1),
 using 5 random inferred trees that appeared 6 or more times in the list of all  2000 inferred trees (Case 2), and  using the ``true" gene trees (Case 3) sampled from the networks.
\label{fig_accuracy_seq}}
\end{figure}

Lastly, we also examined the accuracy of reconstructing a network from the  trees inferred from DNA sequence data using the following setting (see  Methods for details):
\begin{quote}
    -- Generate randomly a network.
    
    -- Sample a gene tree with branch lengths in the network.
    
    -- Simulate DNA evolution to obtain a sequence of 1000 base pairs on the gene tree.
    
    -- Infer a maximal likelihood tree from the simulated sequence. 
\end{quote}
On each random network, we sampled 2000
``true" gene trees and inferred 2000 trees accordingly.

 We examined two networks (Network 3 and Network 4 hereafter,  Supplemental Fig.~S4) on 30 taxa that contain 5 and 6 binary reticulation events, respectively.  Since the inferred trees were noise, we used 5-tree datasets for testing. Inference with more than 5 inferred trees had low accuracy, whereas inference with more than 5 true gene trees had high accuracy.  We ran ALTS on 50 random tree sets for each of the three cases. In the first case, a dataset consists of  5 inferred trees. In the second case, a dataset consists of 5 ``consensus" inferred trees that appeared 6 or more times in the list of inferred trees. Note that a consensus inferred tree is much more likely a true gene tree than a tree that was only inferred once in our experiment.  In the third case, a dataset  
consists of 5 ``true" gene trees.

The results are summarized in 
Figure~\ref{fig_accuracy_seq}. For Network 3, the average dissimilarity score for each inference test was 0.161,  0.079  and 0.043 in Case 1, 2 and 3, respectively. In addition,  ALTS reconstructed Network 3 correctly on 27 out of 50 tree sets in Case 3. The performance of ALTS is similar for the testing on Network 4. These results suggest that accurate inference of gene trees from sequence data is vital for network inference with ALTS. 

\begin{figure*}[t!]
\centering
\includegraphics[scale=0.5]{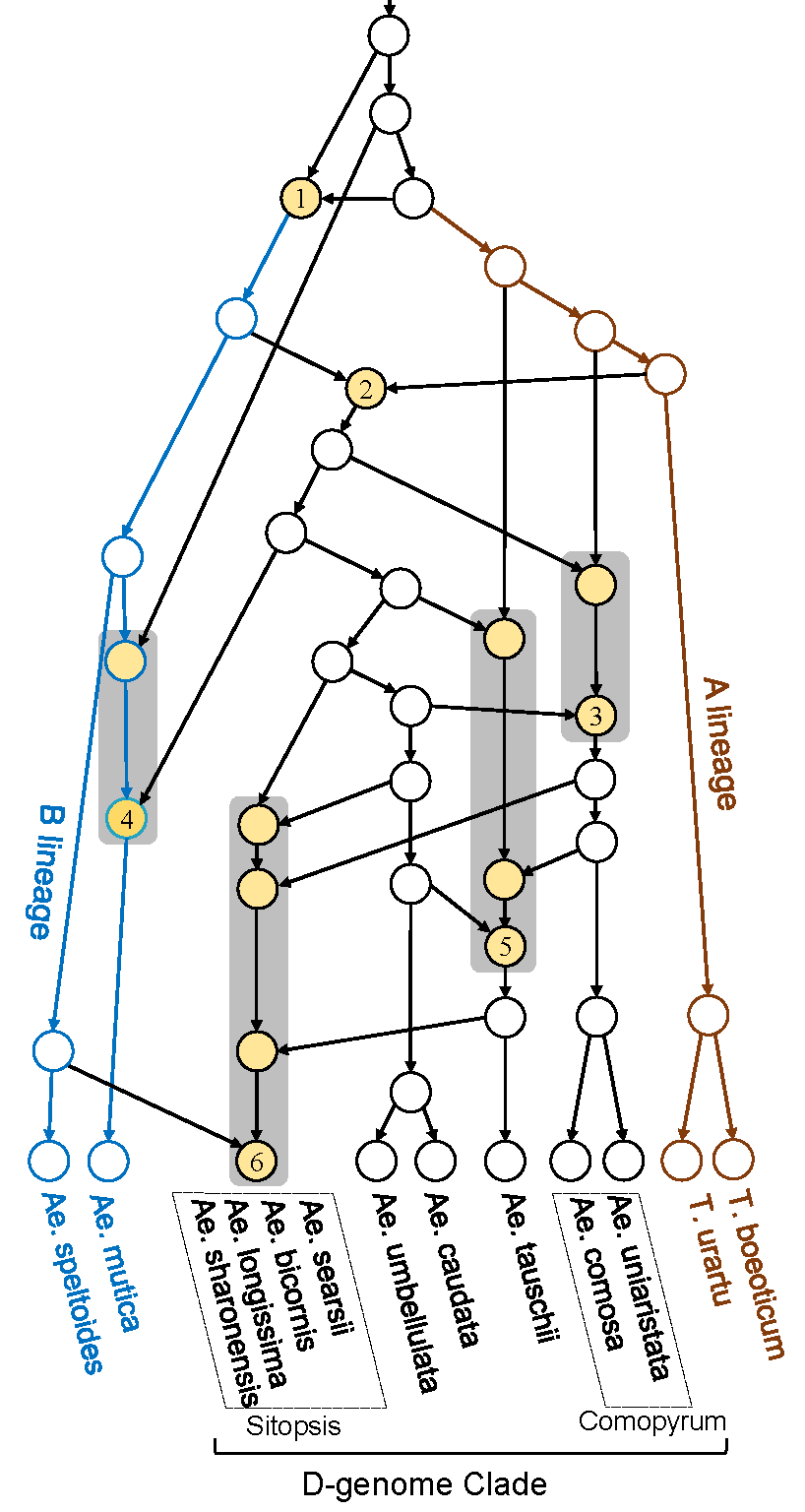}
\caption{\small A phylogenetic network for 13 wheat-related grass species inferred using ALTS. The reticulation nodes are colored in yellow.  The model contains two binary reticulate events (1 and 2) and four event clusters (3 to 6).  
\label{wheat_evol}}
\end{figure*}

\subsection*{A phylogenetic network for 13 wheat-related grass species}

 We also validated our tool by inferring phylogenetic relationships for a set of wheat-related grass species.
Bread wheat ({\it Triticum aestivum}) is a hexaploid species (genome AABBDD) formed through two rounds of hybridization between three diploid progenitors (i.e., {\it T. urartu} of the A subgenome, an unknown species of the B subgenome, and {\it Se. tauschii}  of the D subgenome) (Marcusse et al., 2014; Levy and Feldman, 2022). A recent comprehensive genomic study of Gl\'{e}min {\it et al.} (2019) suggests that hybridizations were pervasive in the evolution of {\it T. urartu}, {\it Se. tauschii} and 11 other grass species. Using a hypothesis testing approach, they detected 6 reliable  and 2 possible  reticulated events. Here, to eliminate the effect of incomplete lineage sorting (ILS) and tree inference errors,
we simplified the 247 gene trees reported in their paper  and selecting 33 of them  for network inference (see Methods section). Using ALTS, we obtained the phylogenetic network depicted in Figure~\ref{wheat_evol}.
The network displays 73 out of 247 simplified gene trees and contains on average 79\% non-trivial node clusters of the remaining trees.

 The network model contains 2 binary reticulate events and 4 clusters of reticulate events. 
Events 2 and 4  and event cluster 6 are consistent with the findings reported by 
Gl\'{e}min {\it et al.} (2019). In particular, Event 2 is the hybridization between A and B lineages that formed the D-subgenome clade (middle, Figure~\ref{wheat_evol}) (Marcusse et al, 2014). The gene flows from an ancestor of {\it Ae. speltoides} in the cluster 6 reveals that the Sitopsis species are more closer to {\it Ae. speltoides} than to {\it Ae. tauschii}, consistent to the cytogenetic analyses reported by Kihara (1954).
Our model also suggests that   complex reticulate events (Cluster 5) occurred between {\it Ae. tauschii} and the  ancestors of {\it Ae. caudata
} and {\it Ae. umbellulata}. 
The complex gene flows in the event cluster 5 has not been reported in literature, but it is compatible with a chloroplast capture model (Fig.~2b, Li et al., 2015). 
Conversely, the two possible reticulate events reported by Gl\'{e}min {\it et al.} (2019) is  not supported by our model. Further verification of these inconsistent interspecific reticulate  events may needs gene order information on the related genomes and additional genomes of D-genome clade. %

\section*{Discussion}
We have presented ALTS. 
It is based on an algorithmic innovation that reduces 
the minimum tree-child network problem to computing the SCS of
the LTSs of the taxa,  obtained from the input trees w.r.t. a predefined ordering on the taxa. 
 ALTS is fast enough to infer a parsimonious tree-child network for a set of 50 trees on 50 taxa in a quarter of an hour on average even if the input trees do not have any non-trivial taxon clusters in common. 
Another contribution is an algorithm for assigning weights to the edges of  the reconstructed tree-child network if the input trees are weighted.  Our work makes network reconstruction more feasible in the study of evolution and phylogenomics.

The accuracy analyses suggest that  50 trees are likely not enough for accurately inferring a phylogenetic network model that has 10 or more reticulation events.
Therefore, a program that can process over hundred trees is definitely wanted. We remark that ALTS can be made even more scalable by distributing the computing tasks for  taxon orderings into a number of processors using distributed computing programming. This is because the computing tasks for different orderings are independent from each other.


 Phylogenetic relationships obtained for the 13 wheat-related grass species and for {\it Hominin} (Supplemental Methods) provide an illustration of good performance of ALTS on empirical data. 
 The analyses show that
  how to eliminate the effects of ILS is important for inference of phylogenetic networks. We will further investigate how to improve the accuracy of ALTS by  incorporating the genomic sequences of the taxa and
  a process of removing ILS events into network inference.

\section*{Methods}

\subsection*{Method for generating random tree datasets}
The simulated tree datasets were generated using an approach appearing in Wu (2010). For each $k\in \{20, 30, 40,  50\}$,   a phylogenetic network on $k$ taxa was first generated by simulating speciation and reticulation events backwards
in time with the weight ratio of reticulation to speciation ratio being set to 3:1. Fifty trees displayed in the networks were then randomly sampled. This process was repeated to generate 2500 trees for each $k$. 

For assessing the accuracy of ALTS for inferring phylogenetic networks from genomic sequences, we generated gene trees with branch lengths from a network by calling a coalescent simulation program  named ms (Hudson 2002). We ordered the speciation and reticulation events in the input network and set the time difference between adjacent evolutionary events to 10 coalescent units. Here, relatively long coalescent time between adjacent evolutionary events was used to reduce the effect of ILS in the simulated gene trees. 

\subsection*{Methods for gene sequence simulation and gene trees inference}

 We used the Seq-Gen program (Rimbaud and Grass 1997) with the GTR  substitution model to generate DNA sequences of 1000 base pairs on a gene tree, where  the scaling factor was set to $0.001$ in order to convert coalescent units to the mutational units for Seq-Gen.  Conversely, we used the RAxML program (Stamatakis 2014) with the GTR model to infer a gene tree from the simulated DNA sequence of 1000 base pairs.   We used an outgroup to root the gene trees inferred by RAxML.

\subsection*{Jaccard distance between two phylogenetic networks}

We measured the dissimilarity between two phylogenetic networks by considering the symmetric difference of the set of taxa clusters in the networks (Huson et al. 2010). Here, a cluster in a network consists of all taxa below a  node in that network.
Precisely, for two phylogenetic networks $N_1$ and $N_2$ over $X$, we use $C(N_i)$ to denote the multiset of clusters appearing in $N_i$ for $i=1, 2$, and define the Jaccard distance between $N_1$ and $N_2$  as  $s(N_1, N_2)=1-|C(N_1)\cap C(N_2)|/|C(N_1)\cup C(N_2)|$. 

\subsection*{Tree data pre-processing for wheat-related grass species}

247 distinct gene trees for 13 wheat-related grass species and 4 outgroup species were downloaded from the evolutionary study of Gl{\'e}min et al. (2019). 
(These trees were inferred from orthologous genes in 47 individual genomes by using RAxML v8.) 
To infer interspecific reticulate events, we simplified the gene trees by using only one individual sequence for each species and removing all 4 outgroup sequences, resulting in 227 distinct trees with 13 leaves. To reduce the effect of ILS and gene tree inferring errors, we further selected 33 gene trees for which either of the following two conditions is true:
(a) it was inferred on two genes; 
(b) every node cluster of it  appears in  $t$ (=20) or more  gene trees.
We used the ratio of the number of trees displayed in a network to its HN to measure its expression capacity. The percentage used in the condition (b) was chosen to control the trade-off between  the size and expression capacity of the network model. %
For $t>18$, the inferred networks had a high HN. For $t>22$, the inferred network displayed a low number of gene trees.  For $t=18, 19, 20, 21, 22$, the HN of the inferred network was 17, 13, 12, 12, 12, whereas the network displayed 90, 71, 71, 71, 63 gene trees, respectively. Since $90/17<71/13<71/12$ and $63/12<71/12$, we selected $20$ as the filtering condition, resulting 33 gene trees.



\section*{Software Availability}

The C source code of ALTS can be found in  Supplemental Source Code. It is also available on https://github.com/LX-Zhang/AAST.

\section*{Competing Interest Statement}

The authors declare no competing interests.

\section*{Acknowledgements} 

We thank Cedric Chauve and Aniket Mane for discussion in the beginning of this project.
We also thank anonymous reviewers for constructive comments on the earlier versions of our manuscript submitted to RECOMB'2023 and Genome Research.
L. Zhang was partly supported by Singapore MOE Tier 1 grant R-146-000-318-114.
Y. Wu was partly supported by U.S. National Science Foundation grants CCF-1718093 and IIS-1909425.

\makeatletter
\renewcommand\@biblabel[1]{}
\makeatother



\clearpage
\newpage 
\begin{singlespace} 
\setcounter{figure}{0}
\renewcommand{\figurename}{Figure}
\renewcommand{\thefigure}{A\arabic{figure}}
\setcounter{page}{1}

\begin{center}
{\large
 {\sc Supplementary Methods} \\
  for \\
 {\bf  A Fast and Scalable Method for
Inferring Phylogenetic Networks\\ from
Trees by Aligning Lineage Taxon Strings} \\
~~\\
   by Zhang et al.
}
\end{center}
\vspace{4em}

A. Correctness of {\sc Algorithm} A \hspace*{14em} - - - - - - - - - - - - Page 2
\\

B. Reduction for a reducible tree set \hspace*{12.5em} - - - - - - - - - - - - Page 11
\\

C. Computing the branch weights of the inferred tree-child network - - - - - - - - - - -  Page 12
\\

D. A phylogenetic network for hominin relationships \hspace*{6.5em} - - - - - - - - - - - Page 14

\newpage 

\section*{A. Correctness of {\sc Algorithm} A}

\subsection*{A1. Concepts and notation}

\subsubsection*{Directed graphs}
A directed graph $G$ consists of a set $V$ of nodes and a set $E$ of directed edges that are ordered pairs of distinct nodes.
Let $e=(u, v)\in E$. We call $e$  an outgoing edge of  $u$ and an incoming edge of $v$.
For a node $v\in V$, its  {\it outdegree} and {\it indegree} are defined as the number of outgoing and incoming edges of $v$, respectively. 

For a graph, {\it subdividing} an edge $(u, v)$ involves replacing it with a directed path from $u$ to $v$ that passes one or more new nodes. Conversely,  an {\it edge contraction} at a node $v$ of indegree one and outdegree one is to remove $v$ and replace the path $u \rightarrow v\rightarrow w$ with an edge $(u, w)$, where $(u,v)$ and $(v, w)$ are the unique incoming and outgoing edge of $v$, respectively. 

\subsubsection*{Phylogenetic networks}
\begin{figure}[t!]
\centering
\includegraphics[scale=0.5]{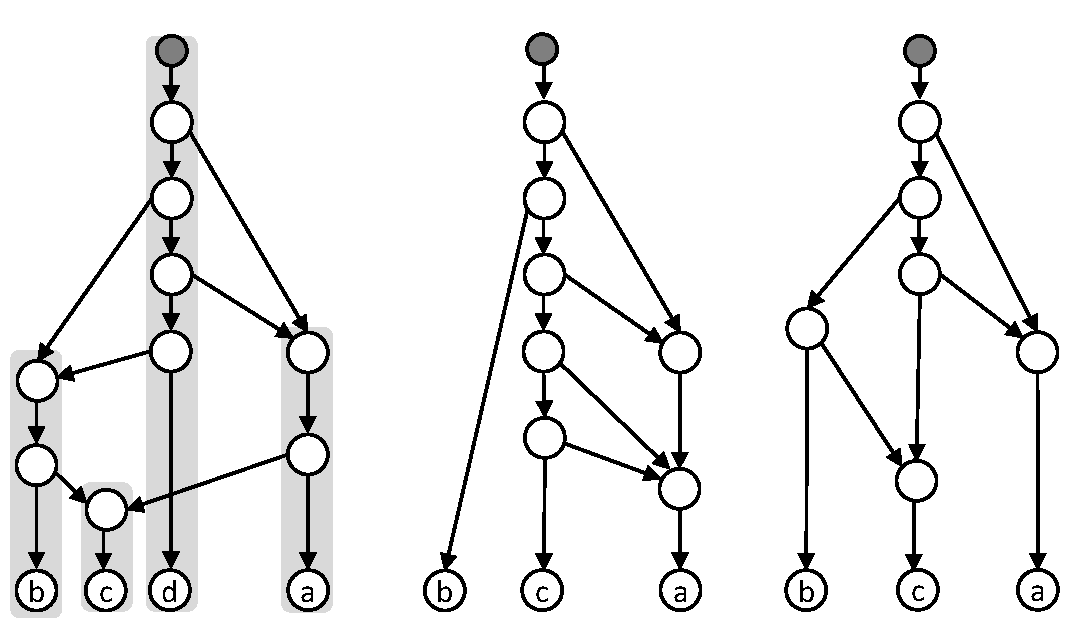}
\caption{\small A binary tree-child network (left) in which there are four tree-node components (shaded grey) and two non-tree-child networks (middle) and  (right). In the middle network,  the child of the top reticulate node is also  reticulate. In the right network, the children of a tree node in the middle are both reticulate. 
\label{FigA1_example} }
\end{figure}

A {\it phylogenetic network} on a set $X$ of taxa is a rooted,  directed acyclic graph in which (i) all the edges are oriented away from the root, which is of indegree 0 and outdegree 1; (ii) the nodes of indegree 1 and outdegree 0, called leaves,  are uniquely labeled with the taxa; and (iii) all the non-root and non-leaf nodes are either tree nodes that are of indegree 1 and outdegree 2 or reticulate nodes that are of indegree more than 1 and outdegree 1. Reticulate nodes represent  evolutionary reticulation events. A phylogenetic network is said to be {\it binary} if the indegree of every reticulate node is exactly 2 (Figure~\ref{FigA1_example}).

Let $N$ be a 
phylogenetic network.  We use ${\cal V}(N)$ and ${\cal E}(N)$
to denote the node and edge set of $N$, respectively. We also use ${\cal R}(N)$ to denote the set of reticulate nodes, and use ${\cal T}(N)$  to denote the set of all non-reticulate nodes, including the root, tree nodes and leaves.
Let  $u, v\in {\cal V}(N)$. The node $v$ is a {\it child}  of $u$ if $(u, v)$ is an edge;  $v$ is a {\it descendant} of $u$ if there is a directed path from $u$ to $v$. If $v$ is a descendant of $u$,  $v$ is said to be {\it below} $u$.

A phylogenetic network $N$ is a  {\it tree-child network} if every non-leaf node has a child that is not reticulate. Equivalently, $N$ is a tree-child network if and only if for every non-leaf node, there is a path from that node to some leaf that passes only tree nodes.
Figure~\ref{FigA1_example} presents  a binary tree-child network (left) and two non-tree-child networks.  %

Consider a tree-child network $N$ with $k$ reticulate nodes.
Let the root be $r_0$ and  let the reticulate nodes be  $r_1,
r_2, \cdots, r_k$. After the removal of the incoming edges of every $r_i$, $N$ becomes the union of $k+1$ subtrees, which are rooted at $r_0, r_1, \cdots, r_{k}$, respectively,  and have network leaves as their leaves (see Figure~\ref{FigA1_example}). These subtrees are called the {\it tree-node components} of $N$.
Tree-node decomposition is a useful technique in the study of phylogenetic networks. 

\subsubsection*{Phylogenetic trees}

A phylogenetic tree on $X$ is a phylogenetic network with no reticulate nodes.  In fact, a tree is a tree-child network.
Let $T$ be a phylogenetic tree on $X$ and $u\in V(T)$. The {\it node cluster} of $u$, denoted as $C(u)$, is  the subset of taxa that are represented by the leaves below $u$. Clearly,  $C(u)\cap C(v) \in \{C(u), C(v), \emptyset\}$ for any two nodes $u$ and $v$.
The node $u$ and its descendants induce a unique subtree on 
$C(u)$.  We use $T_u$ or $T(C(u))$ to denote the subtree.

Let $S$ be a set of binary phylogenetic trees on $X$. A {\it common cluster} of $S$ is a  subset of $X$  that is a node cluster in every tree of $S$. Obviously, each single taxon is  common cluster of $S$, and so is $X$. Any other common clusters of $S$ are called {\it non-trivial common clusters}.
$S$ is a {\it reducible} tree set if there is a non-trivial common cluster for $S$,  and it is {\it irreducible} otherwise.    A non-trivial common cluster $C$ of
$S$ is {\it maximal} if any subset $C'$ such that $C\subset C'\subset X$ is not a common cluster of $S$.  Clearly, for any two maximal common cluster $C_1$ and $C_2$ of $S$, $C_1\cap C_2=\emptyset$;
and any non-trivial common cluster $X'$ of $S$ must be contained in a unique maximal cluster of $S$ if $X'$ is not maximal.

\subsubsection*{Tree display and network inference problems}

Let $T$ be a binary phylogenetic tree on $X$ and let $N$ be a  tree-child network with $k$ reticulate nodes on $X$. $T$ is {\it displayed} by
$N$ if $T$ can be obtained from $N$ by applying edge contraction from  $N$ after the removal of all but one incoming edge for each reticulation node  (Figure~\ref{display}).  For any set of binary phylogenetic trees over $X$, there is always a tree-child network that displays all the trees  \cite{linz2019attaching}. However,  such a solution network may not be binary.
%
%

Let $P$ by a phylogenetic network.
Its {\it reticulate number} is defined as the number of reticulate nodes. Its {\it HN}, denoted as
$H(P)$,  is defined as the sum over all the reticulate nodes of the difference between the indegree and the outdegree of that reticulate node. If $P$ is binary, $H(P)$ is equal to the reticulate number.  Here, we studied the following {\it minimum tree-child network inference} problem:
\\
\\
%
\begin{tabular}{l}
{\bf Input}: A set of phylogenetic trees on  $X$.\\
{\bf Output}:  A parsimonious  tree-child network $P$  on $X$ (with the smallest $H(P)$) that\\ displays all input trees.
\end{tabular}

\subsubsection*{The SCS problem}

Let $s$ and $t$ be two sequences in an alphabet. The sequence $s$ is said to be a supersequence of $t$ if  $t$ can be obtained from $s$ by the deletion of one or more letters. The {\it SCS problem} is, given a set of sequences, to find the shortest sequence that is a supersequence of every given sequence.  

 The SCS problem can be solved in a quadratic time for two sequences. However,  it is NP-hard in general. 


\subsubsection*{Total ordering}

Let $X$ be a set of taxa. A (total) ordering $R$ on $X$ is a binary relation on $X$ such that (i) $R$ is anti-symmetric, i.e. if $x_1Rx_2$, then $x_2\not\hspace*{-0.4em}R~x_1$. (ii) $R$ is transitive, i.e., 
if $x_1 R x_2$ and $x_2 R x_3$, then $x_1Rx_3$.
(iii) For any $x_1, x_2$, $x_1Rx_2$ or $x_2Rx_1$.
For convention, we write $x<_R y$ if $x$ is related $y$ under $R$ or even $x<y$ if $R$ is clear.

Any non-empty subset $X'$ of $X$ whose elements are ordered according to $R$ has a unique minimum (resp. maximum) element. We use $\min_R X'$ (resp. $\max_R X'$) to denote the minimum (resp. maximum) element of $X'$.

Let $X=\{x_1, x_2, \cdots, x_n\}$. We use  $\pi=\pi_1\pi_2\cdots \pi_n$ on $\{1, 2, .., n\}$ to denote the following ordering:
$$ x_{\pi_1} < x_{\pi_2} < \cdots < x_{\pi_n}.$$

  


\subsection*{A2. Proof of Propositions}

We use the following algorithm to derive another representation of a phylogenetic tree on $|X|$ given an ordering on $X$.
\\
\\
\begin{tabular}{l}
\hline
{\sc Labeling} \\
\\
{\bf Input}  A tree $T$ on $X$ and an ordering $\pi$ of $X$\\
\\
1. Label the degree-1 root  of $T$ by $\min_\pi(X)$. \\
2. Label each internal node $u$ with
 two children $v$ and $w$ with \\
\hspace*{1.5em}$\max_{\pi}\{\min_{\pi}(C(v)), \min_{\pi}(C(w))\}$, where $C(v)$ consists of all taxa below $v$ in $T$. \\
\hline
\end{tabular}
\\
\\

 For each taxon $\tau$,  a unique internal node $w$  is labeled with $\tau$ by applying the {\sc Labeling} algorithm. The node $w$ is an ancestor of the leaf $\tau$.  Let $Z_{w\tau}$ be the directed path from $w$ to the leaf $\tau$ in the tree. The
sequence of the labels of the nodes appearing between $w$ and the leaf  in the path $Z_{w\tau}$ is called
{\it the lineage taxon string} (LTS) of  $\tau$.
\\
\\
\noindent {\bf Proposition 1}.
Let $\pi$ be an ordering of $X$, $|X|>1$. For a phylogenetic tree $T$ on $X$, the LTS $s_{\pi}(t)$ of each taxon $t$ obtained w.r..t $\pi$ by applying the {\sc Labeling} algorithm in $T$ has the following properties:

 (i) $s_\pi(\pi_1)$ is always not empty; 
    
  (ii)   $s_{\pi}(\pi_n)$ is always empty;

(iii) for each $1<i\leq n$,    $\pi_i$ appears in the LTS of $\pi_j$ for a unique $j$ such that $j<i$;

(iv) the smallest taxon $\pi_1$ does not appear in any LTS.
\\

\noindent{\bf Proof.} 
Let the degree-1 root of $T$ be $\rho$. Let the ancestors of Leaf $\pi_1$ be:
$$\rho=u_0, u_1, u_2, \cdots, u_k$$
and $u_{k+1}=\pi_1$, 
where $u_{i}$ is the parent of $u_{i+1}$ for
$0\leq i\leq k$. Recall that each non-leaf, non-root node has two children.  We let $u'_{i+1}$ be another child of 
$u_i$ for $0\leq i\leq k$. 

(i) Since $|X|>1$, $k\geq 1$. Clearly, 
$\min_\pi C(u_i)=\pi_1$ for each $i\leq k$. Since $\pi_1$ is the smallest taxon, 
in Step 2 of the 
{\sc Labeling} algorithm, $u_i$ is
labeled with $\max_\pi \{\min_\pi(u_{i+1}),  \min_\pi(u'_{i+1})\}=\min_\pi(u'_{i+1})$ for
$i=1, 2,  \cdots, k$. Therefore, that $k\geq 1$ implies that $s_\pi(\pi_1)$ contains at least one taxon.

(ii) Let the parent and sibling of Leaf  $\pi_n$ be $v$ and $v'$. In Step 2 of the 
{\sc Labeling} algorithm, $v$ is
labeled with $\max_\pi \{\min_\pi(v'),  \pi_n\}=\pi_n$. Since there is no node between $v$ and Leaf $\pi_n$,  $s_\pi(\pi_n)$ is empty.

(iii) and (iv) We prove the statement by mathematical induction. If $|X|=2$, clearly, the root $\rho_T$ is labeled with $\pi_1$ and the other internal node is labeled with $\pi_2$. In this case,  $s_\pi(1)$ contains only $\pi_2$ and
$s_\pi(2)$ is empty. Thus, the fact is true.

For $|X|>2$, from the proof of Part (i), we have that
$u_i$ is labeled with the minimum taxon appearing in $C(u'_{i+1})$ for $i=1, 2, \cdots, k$. Moreover, the internal nodes in each subtree $T'_i$ rooted at $u'_{i}$ are labeled with the taxa of
$C(u'_{i})\setminus \{\;\min_\pi C(u'_i)\;\}$ according to the algorithm. Since each $T'_i$ is a proper subtree of $T_i$, by induction, the fact holds. $\square$
\\
\\
\noindent {\bf Remark}. 
The  LTSs of the taxa obtained according to an ordering on $X$ determine a unique phylogenetic tree $T$.
This can be generalized to an algorithm to reconstruct a tree-child network using the
 LTSs of taxa.
\\
\\
 \begin{tabular}{l}
 \hline
 {\sc Tree-Child Network Construction} \\
 \\
  1. ({\bf Vertical edges}) For each $\beta_i$,  define a path $P_i$ with $\vert \beta_i\vert +2$ nodes: \\
  \hspace*{3em} $h_i, v_{i1}, v_{i2}, \cdots, v_{i\vert\beta_i\vert}, \ell_{\pi_i}$, where $\beta_n$ is the empty sequence.\\
  2.  ({\bf Left--right edges}) 
  Arrange the $n$ paths 
  from left to right as $P_1, P_2, \cdots, P_n$.  If the\\
  \hspace*{2em}$m$-th letter of $\beta_i$ is $\pi_j$, we add an edge 
  $(v_{im}, h_{j})$ for each $m$ and each $i$. \\
  3.  Contract each $h_i$ ($i>1$) if it is of indegree 1 and outdegree 1.\\
    \hline
 \end{tabular}
 \\
 \\
 \\
\noindent {\bf Proposition 2}. Let $T_i$ ($1\leq i\leq k$) be $k$ trees on $X$ such that $\vert X\vert =n$ and $\pi$ be an ordering on $X$. Let $\alpha_{ij}=\beta_{T_i, \pi}(\pi_j)$,  the LTS of $\pi_j$  with respect to $\pi$ in $T_i$ for each $j$, $1\leq j\leq n -1$. If $\beta_j$ is a common supersequence of 
$\alpha_{1j}, \alpha_{2j}, \cdots, \alpha_{kj}$ for each $j$, the  {\sc Tree-Child Network Construction} algorithm  outputs a tree-child network that displays the $k$ trees.
\\
\\
\noindent {\bf Proof.} 
Let $N$ be the directed network constructed by applying  the algorithm to $\beta_1, \beta_2, \cdots, \beta_k$.
First, $N$ is acyclic due to the two facts:
(i) the edges of each path $P_i$ are oriented downwards, and (ii) the so-called left--right edges $(u, v)$ are oriented from a node $u$ in a path defined for $\pi_i$ to a node $v$ in a path defined for
$\pi_j$ such that $i <j$. 

Second, $N$ is tree-child. This is because all the nodes of each $P_i$  are tree nodes except $h_i$ for each $i>1$ (see Figure 3 in main text).  The node $h_1$ is the network root. For $i>1$, $h_i$ may or may not be a reticulation  node. Therefore, every non-leaf node has a child that is not reticulate. 

Lastly, we prove that $T_i$ is displayed by $N$ as follows. By assumption, $\beta_j$ is a supersequence of $\{\alpha_{ij} \;\vert\; i=1, 2, \cdots, k\}$ for each $j=1, 2, \cdots, n-1$.
Following the notation used in the {\sc Tree-Child Network Construction} algorithm,
we let:
$$\beta_j=\beta_{j1}\beta_{j2}\cdots \beta_{jt_j},\;\;t_j\geq 1,$$
where $t_j$ is the length of $\beta_j$.
Since $\alpha_{ij}$ is a subsequence of $\beta_j$, there is an increasing subsequence 
$1\leq m_1 < m_2 <\cdots < m_{\ell_j} \leq t_j$
such that
$$ \alpha_{ij}=\beta_{im_1}
\beta_{im_2}\cdots \beta_{im_{\ell_j}}$$
and $\ell_j=|\alpha_{ij}|\geq 1$.

According to Step 1 of the algorithm, in  $N$, each taxon $\beta_{jx}$ of $\beta_j$ corresponds one-to-one a node $v_{jx}$ in the path $P_j$; and
there is a (left-right) edge from $v_{jx}$ to the first node $h_{y(x)}$
of the path $P_{y(x)}$ that ends with 
the taxon $\pi_{y(x)}=\beta_{jx}$, where $y(x)\geq j$. 

Conversely, after removing the edge 
$(v_{jx}, h_{y(x)})$ for each $x\neq m_1, m_2, \cdots, m_{\ell_j}$, we obtain a subtree $T'_i$ of $N$. This is because each taxon $\pi_t$ appears exactly once in $\alpha_{i1}, \alpha_{i2}, \cdots, \alpha_{i(n-1)}$ and thus the node $h_t$ is of indegree 1 in the resulting subgraph, where $t=2, 3, \cdots, n$. 
It is not hard to see that after contracting  degree-2 nodes of $T'_i$, the resulting subtree $T''_i$ has the same LTS as $T_i$ for each $\pi_j$. Thus $T''_i$ is equal to $T_i$.  $\square$
\\
\\
\noindent {\bf Definition 1}. Let 
$P$ be a phylogenetic network on
$X$, where $|X|>1$ and $\pi$  be an ordering on $X$.  
$P$ is said to be {\it compatible} with $\pi$ if
for each reticulate edge $(s, r)$ of $P$, the minimum  taxon below $s$ in the tree-node component $C_s$ is less than the minimum taxon in the tree-node component $C_r$.
\\
\\
{\bf Remark.} For a tree-child network $P$, we can construct a compatible ordering $\pi$ as follows. We first compute a topological sorting on the vertices of $P$.  Assume the reticulate nodes and the network root $\rho$ appear in the sorted list as: $r_0=\rho, r_1, r_2, \cdots, r_k$. We construct a desired ordering by listing the taxa in the tree-node component $C_{r_i}$ before the taxa in the tree-node component $C_{r_{i+1}}$ for every $i\leq k-1$.
\\
\\
Let $\pi$ be an ordering on $X$
and  $P$ be a tree-child network on $X$  that is compatible with
$\pi$. The compatibility property implies that the smallest taxon is in the tree-node component $C_{\rho}$ that is rooted at the network root $\rho$.  We use the following generalized {\sc Labelling} algorithm to label all the tree nodes of $P$, which is identical to {\sc Labelling} when $P$ is a phylogenetic tree.
\\
\\
\begin{tabular}{l}
\hline 
{\sc Generalized Labelling}\\
\\
 {\bf S1}: For every reticulate node  $r$,  label all parents of  $r$ with the smallest taxon in \\ 
 \hspace*{1em} the tree-node component $C_r$. 
 Similarly, the network root $\rho$ is labeled with\\
 \hspace*{1em} the smallest taxon in  $C_\rho$.\\
 {\bf S2}: For each tree node $z$ that is not a parent of any reticulate node, label $x$ with \\
 \hspace*{1em} $\max_{\pi} (\min_\pi(C(x)), \min_\pi(C(y))$,
 where  $x$ and $y$ are the two children of $z$, and \\
 \hspace*{1em} $C(x)$ and $C(y)$ are the set of taxa  below $x$ and $y$ in the tree-node component\\
 \hspace*{1em} where they belong to.\\
\hline
\end{tabular}
\\
\\
\\
\noindent {\bf Proposition 3}. Let $T_1, T_2, \cdots, T_k$ be $k$ trees on $X$ and $P$ be a tree-child network on $X$ with the smallest $H(P)$,  compared with those displaying all $T_i$. For any ordering $\Pi$ of $X$ such that $P$ is compatible with it, if we label the tree nodes of $P$ using the {\sc Generalized Labelling} algorithm, the LTS $\beta_{P, \Pi}(x)$ obtained for each taxon $x$ is a shortest common supersequence of 
 $\{\beta_{T_i, \Pi}(x) \;\vert\; i=1, 2, \cdots, k\}$. 
Moreover, applying the {\sc Tree-child Construction} algorithm to the obtained
supersequences $\beta_{P, \Pi}(x)$ produces the 
same network as $P$.
\\
\\
The proof of Proposition 3 is divided into several lemmas.
\\
\\
\noindent {\bf Lemma 1}.~ 
Let $\pi$ be an ordering on $X$ and let 
$T_1, T_2, \cdots, T_k$ be $k$ phylogenetic trees on $X$. For each $x\in X$ and each $T_i$, we use $\beta_x(T_i, \pi)$ to denote the LTS of $x$ obtained w.r.t. $\pi$ using the {\sc Labeling} algorithm in $T_i$. Assume $\beta_x$ is a common supersequence of $\{\beta_x(T_1, \pi),\beta_x(T_2, \pi), \cdots, \beta_x(T_k, \pi)\}$ for each $x\in X$.
For the tree-child network $P$ constructed from  $\{\beta_x \;\vert \; x\in X\}$ by using the {\sc Tree-Child Network Construction} algorithm, $H(P)=\sum_{x\in X}|\beta_x|-|X|+1.$
\\
\\
\noindent {\bf Proof.} Since 
only the first node $h_i$ of each path can be a reticulate node and that each node in the middle of each path is a parent of some $h_i$, 
$H(P)=\sum^{|X|}_{i=2} (d_{in}(h_i)-1)= \sum_{x\in X}|\beta_x|-|X|+1$, where
$d_{in}(h_i)$ is the indegree of $h_i$. $\square$
\\
\\
\begin{figure}[!b]
    \centering
    \includegraphics[scale=0.6]{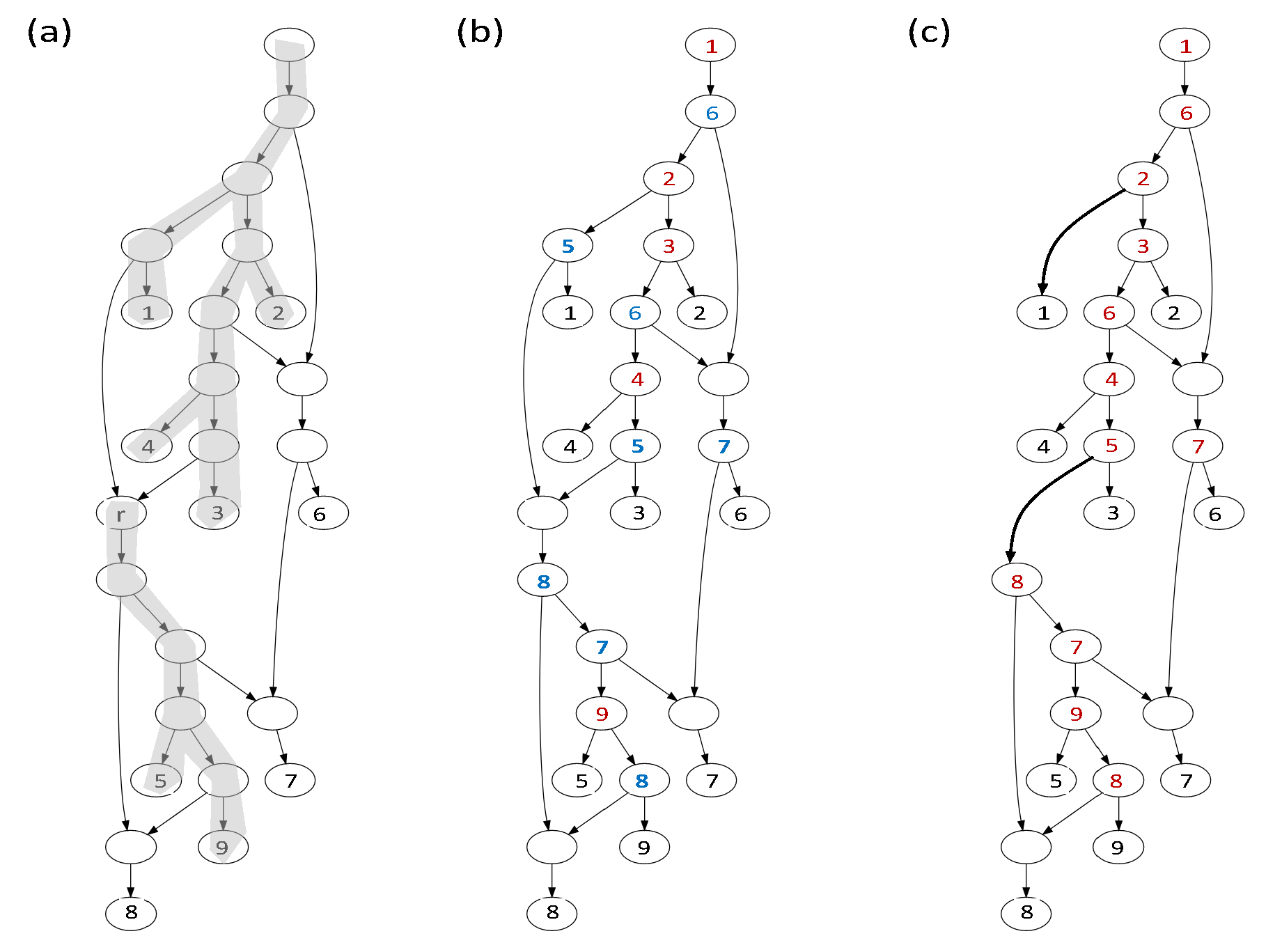}
    \caption{Illustration of the {\sc Generalized Labelling} algorithm and the proof of Lemma 3. 
    (a) A tree-child network on the taxa from 1 to 9, which has two tree-node components each containing at least two taxa. (b) Labelling all the tree nodes in a tree-child network using the increasing order of taxa:
    $i<i+1, i=1, 2, ..., 8$, which is compatible.  The labels of the parents of a reticulation node are in blue; while the labels of other tree-nodes are in red.
    (c) the resulting network after the removal of the left incoming edge of the reticulation node $r$, in which the tree-nodes are labeled identically if the same ordering is used.}
    \label{fig:A2}
\end{figure}

\noindent {\bf Lemma 2.}  Let $C$ be a tree-node component of $P$ and let it contain $t$ taxa $x_1, x_2, \cdots, x_t$ in $P$. 
All $t-1$ tree nodes that are not a parent of any reticulate node are uniquely labeled with some $x_j \neq 
\min_{\pi}\{x_i \;\vert\; 1\leq i\leq t\}$ (red labels in Figure~\ref{fig:A2}b).
\\
\\
\noindent {\bf Proof.} This can be proved using the same mathematical induction as in Prop.~1.iii.
 $\square$ 
 \\
\\
\noindent {\bf Definition 2.}
Let $\pi$ be an ordering on $X$ and $N$ be a tree-child network on $X$ that is compatible with $\pi$. 
Assume the tree nodes of $N$ are labeled by using the 
{\sc Generalized Labelling} algorithm.
The LTS 
of a taxon $x$ obtained according to 
$\pi$ is defined to be the sequence of the labels of the $x$'s ancestors that are  a tree node in  $C_x$,  if $x$ is the smallest taxon in $C$;  it is the sequence of the labels of the $x$'s  ancestors  that are a  tree-node below the unique tree node labeled with $x$ in $C_x$ otherwise.
The LTS of $x$ obtained in this way is denoted by $\beta_{N,\pi}(x)$.
\\
\\
\noindent {\bf Definition 3}. Let $P$ be a tree-child network on $X$ and let $(s, r)$ be a reticulate edge. $P-(s, r)$ is defined to be the tree-child network obtained through the removal of $(s, r)$ and contraction of $s$ (and also $r$ if $r$
is of indegree 2 in $N$).
\\
\\
\noindent {\bf Lemma 3}. Let $\pi$ be an ordering on $X$ and $P$ be a tree-child network on $X$ such that $H(P)\geq 1$ and $P$ is compatible with $\pi$. For any reticulate node $r$ and each parent $s$ of $r$, the tree-child network $P-(s, r)$ has the following properties:
\begin{enumerate}
      \item $P-(s, r)$ is also compatible with $\pi$;
    \item For each taxon $x$,  $\beta_{P, \pi}(x)$ is a supersequence of $\beta_{P-(s, r),\pi}(x)$.
 \end{enumerate}  
 \noindent {\bf Proof.} These properties are illustrated in Figure~\ref{fig:A2}.    Let $(s, r)$ be a reticulate edge. We have that $s$ is a tree node,  and $r$ is a reticulate node. 
 Recall that $C_N(z)$  denotes the tree-node component containing $z$ for each node $z$ and  for $N=P$, or $P-(s, r)$.  We consider the two cases.
 \\
 
 \noindent {\bf Case 1.} The $r$ is of indegree 3 or more. 
 
 In this case, after $(s, r)$ is removed, $s$ will be contracted and all the other nodes remains the same in $P-(s, r)$. Moreover, $P-(s, r)$ has the same tree-nodes components as $P$ and also has the same labelling as $P$. 
 For any reticulate edge
 $(s', r')$, $C_{P-(s, r)}(s') =  C_{P}(s')$ and $C_{P-(s, r)}(r')= C_{P}(r')$. As such, the constraint is also satisfied for $(s', r')$ in $P-(s, r)$. Therefore, the first fact holds.
 
Let  $x$ be a taxon.  If $\beta_{P, \pi}(x)$ contains the label $y$ of $s$, say
 $\beta_{P, \pi}(x)=\beta_1y\beta_2$, then,
 $\beta_{P-(s, r), \pi}(x)=\beta_1\beta_2$. If $\beta_{P, \pi}(x)$ does not contain the label of $s$, $\beta_{P-(s, r), \pi}(x)=\beta_{P, \pi}(x)$. This concludes that $\beta_{P, \pi}(x)$ is a supersequence of $\beta_{P-(s, r), \pi}(x)$. Therefore the second fact is true.
 \\
 
 \noindent {\bf Case 2.} The $r$ is of indegree 2.  
 
 This case is illustrated in Figure~\ref{fig:A2}b.
 Let $s'$ be another parent of $r$.  After 
 $(s, r)$ is removed, the $r$ becomes a node of indegree 1 and outdegree 1 and thus is contracted, together with $s$. All the other nodes remains in $P-(s, r)$.  Therefore, $s'$ becomes a tree node 
  in $P-(s, r)$. The tree-node component $C_{P-(s, r)}(s')$ is the merge of 
  $C_P(s')$ and $C_{P}(r)$.  Assume  $(s'', r')$ be a reticulate edge
 of $P-(s, r)$. 
 
 If 
 $C_{P-(s, r)}(s'') \neq  C_{P-(s, r)}(s')$
 and $C_{P-(s, r)}(r') \neq C_{P-(s, r)}(s')$, then, $C_{P-(s, r)}(s'')=C_{P}(s'')$ and 
 $C_{P-(s, r)}(r')=C_{P}(r')$. The constraint is satisfied for $(s'', r')$.
 
 If $C_{P-(s, r)}(s'') \neq  C_{P-(s, r)}(s')$ and $C_{P-(s, r)}(r') = C_{P-(s, r)}(s')$, the constraint is satisfied  for
 $s'', r'$ because of the fact that 
 $\min_\pi C_{P-(s, r)}(r')=\min_\pi C_{P}(r')$.
 
 If $C_{P-(s, r)}(s'') =  C_{P-(s, r)}(s')$ and $C_{P-(s, r)}(r') \neq C_{P-(s, r)}(s')$, then the minimum taxon below $s''$ in $C_{P-(s, r)}(s'')$ is equal to 
 that in $C_{P}(s'')$, the constraint is satisfied for $(s'', r')$.
 
  We have proved the first statement.   We prove the second statement as follows.
  To this end, we use $c_P(r)$ to denote the unique child of $r$ in $P$. 
  
  Recall that after $(s, r)$ was removed, $s$ and $r$ were contracted to obtain $P-(s, r)$. Note that 
 in $P-(s, r)$, $s'$ becomes the parent 
 of $c_P(r)$. Since $P$ is compatible with 
 $\pi$, the minimum taxon $y$ below $c_P(r)$ is larger than the minimum taxon below $s'$ in $\pi$. This implies that $s'$ is labeled with $y$, as $s'$ is not a parent of any reticulate node in $P-(s, r)$. Therefore, 
 for any taxon $x\in X$, if $\beta_{P, \pi}(x)$ contains the label $y$ of $s$,  say $\beta_{P, \pi}(x)=\beta_1y\beta_2$, then, 
 $\beta_{P-(s, r), \pi}(x)=\beta_1\beta_2$. If $\beta_{P, \pi}(x)$ does not contain the label of $s$, $\beta_{P-(s, r), \pi}(x)=\beta_{P, \pi}(x)$. This concludes that $\beta_{P, \pi}(x)$ is a supersequence of $\beta_{P-(s, r), \pi}(x)$ for each $x\in X$. $\square$
 \\
\\
 \noindent {\bf Proof of Proposition 3}. 
 Let $P$ be a tree-child network on $X$
 with the smallest $H(P)$, compared with those displaying all $T_i$. For each $i$, 
 $T_i$ can be obtained from $P$ by deleting all but one incoming edge for each reticulate node. For convention, we assume that all removed reticulate edges are 
 $(s_j, r_j)$, $1\leq j\leq H(P).$  
Let $x$ be a taxon.  By Lemma 3, $\beta_{P, \Pi}(x)$ is a supersequence of $\beta_{P-(s_1, r_1), \Pi}(x)$ and 
$\beta_{P-\sum^{j}_{t=1}(s_t, r_t), \Pi}(x)$ 
is a supersequence of $\beta_{P- \sum^{j+1}_{t=1}(s_t, r_t), \Pi}(x)$ for each $j=1, .., H(P)-1$. 
Therefore, for any $x$, $\beta_{P,\Pi}(x)$ is a supersequence of $\beta_{T_i, \pi}(x)$ for each $T_i$, as 
$T_i=P- \sum^{H(P)}_{j=1}(s_j, r_j)$.

Let $P$ contain $m$ reticulate nodes. $P$ has $m+1$ tree-node components. 
In a tree-node component $C$, there are 
$|X(C)|-1$ tree nodes that are not the parents of any reticulation nodes, where 
$X(C)$ is the set of taxa in $C$. Hence
\begin{eqnarray*}
  && \sum_{x\in X}|\beta_{P, \Pi}(x)|\\
  &=& \sum_{C} (|X(C)|-1) + \sum_{r\in {\cal R}(P)} d_{in}(r)\\
  &=& |X|-(m+1) + H(P) + m\\
  &=& |X| -1 +H(P).
\end{eqnarray*}
This implies that $H(P)=\sum_{x\in X}|\beta_{P, \Pi}(x)|-|X|+1.$ 

Assume $\beta_{P, \Pi}(x)$ is not a shortest supersequence of $\beta_{T_i,\Pi}(x)$ ($i=1, 2, \cdots, k$) for some $x$. Let $\beta_x$ be a shortest supersequence of $\beta_{T_i,\Pi}(x)$ ($i=1, 2, \cdots, k$). Then, 
$|\beta_x|< |\beta_{P, \Pi}(x)|$.  By Lemma 1, we can use the {\sc Tree-Child Network Construction} algorithm to obtain a tree-child network with the HN smaller than $H(P)$, a contradiction. 

 It is obvious that the we obtain $P$ if the {\sc Tree-Child Network Construction} algorithm is applied to the LTSs $\beta_{P, \Pi}(x)$ of the taxa $x$. 
$\square$

\section*{B. Reduction for a reducible tree set }

A set of multiple trees is reducible if there is a non-trivial node cluster that appears in every tree and is irreducible otherwise. 
ON way for improving the scalability is
to decompose the input tree set into irreducible sets of trees if the input trees are reducible.

\begin{figure}[t!]
\centering
\includegraphics[scale=0.8]{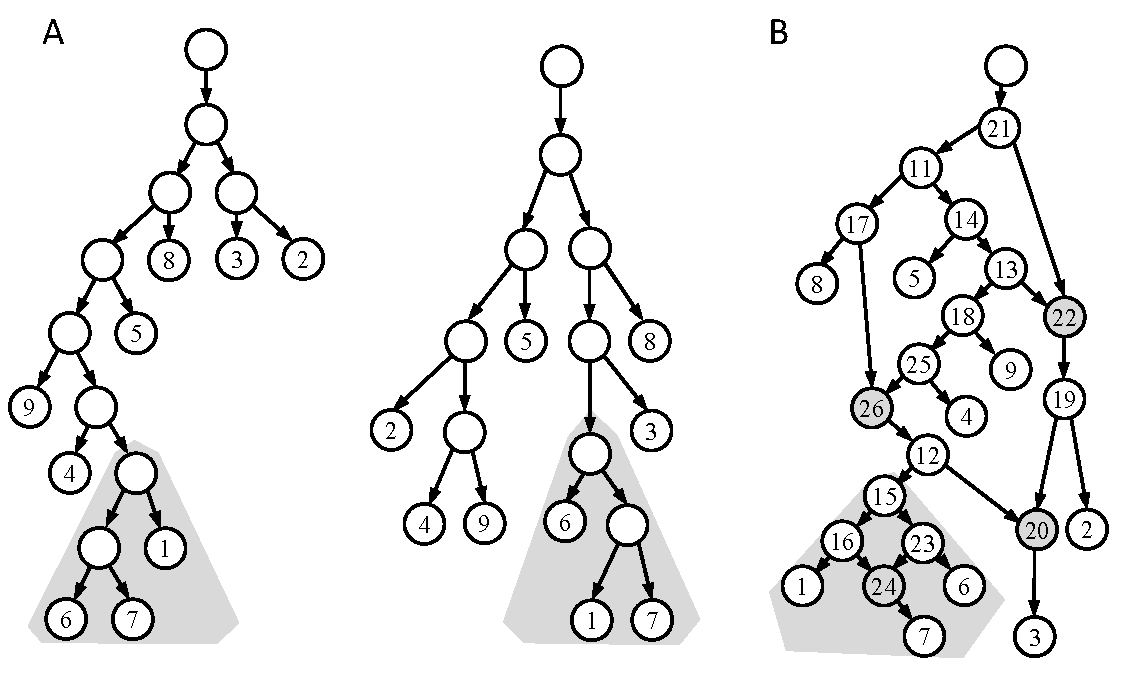}
\caption{A. Two input trees over taxa 1–9 that contain an identical node cluster:
(1, 6, 7). B. A tree-child network that display both input trees, which
is a merge of two tree-child networks.
\label{FigA3}}
\end{figure}

Let $S$ be a reducible set of $k$  trees on $X$,  which are ordered as: $\left<T_1, T_2, \cdots T_k\right>$.
We assume that $C_1, C_2, \cdots, C_t$ are all the maximal common clusters of $S$. We introduce $t$ new taxa $y_i$ and let $Y=\{y_1, y_2, \cdots, y_t\}$. 
By replacing $T_i(C_j)$ with $y_j$ in $T_i$ for each $i$ and $j$,  
we obtain a set $S'$ of $k$  trees $T'_i$ on 
$Y\cup \left[X\setminus \left(\cup^{t}_{i=1}C_i\right)\right]$.
In this way, we decompose $S$ into an irreducible tree set 
$S'=\left< T'_1, T'_2, \cdots, T'_k\right>$ and $t$ ordered sets of trees
$S'_i=\left<T_1(C_i), T_2(C_i), \cdots, T_k(C_i) \right>$, $1\leq i\leq t$. 
Combining the tree-child networks constructed from $S'$ and all of $S'_i$ gives tree-child networks that display all the trees of $S$, as shown in Figure~\ref{FigA3}.

\section*{C. Computing the branch weights of the inferred tree-child network}
%
%
%
%
%
%
%

A phylogenetic network is weighted if every branch has a non-negative value, which represents time or other evolutionary measures. 
A weighted phylogenetic tree $T$ is said to be displayed in a weighted network $N$ if the tree is displayed in the network when  the branch weights are ignored. For  a display  $T'$ of $T$ in $N$,   its {\it fitness score}
$||T-T'||_2$  is defined as $\sqrt{\sum_{e\in E(T)}|w_T(e)-w_{T'}(P(u', v'))|^2}$, where $w_T(e)$ is the weight of $e=(u, v)$ in $T$ and 
$w_{T'}(P(u', v'))$ is the weight of the unique path between the images $u'$ and $v'$ of $u$ and $v$ under the display mapping, respectively.  

Recall that a tree can be displayed multiple times in a network. 
The 
{\it score} of the display of $T$ in $N$ is the smallest fitness score which a display of $T$ in $N$ can have, denoted $d(T, N)$. If $d(T, N)=0$, we say that $N$ perfectly displays $T$.


If the input trees are weighted, 
we will first compute  tree-child networks that each display all the trees.
We then use branch weights of trees and the information on how the trees are displayed in a tree-child network to compute the weights of the network branches.



We model the branch weight assignment problem as an optimization problem with the following assumption on the inferred tree-child network $N$ that displays all the trees:
\begin{quote}
  For any reticulate edge $e$, 
  the tree-child network $P-e$ obtained after removal of $e$ fails to display one input tree at least.
\end{quote}
By ordering the edges of $N$ on $X$, we may assume  
$$E(N)=\{e_1, e_2, \cdots, e_m\}.$$
Let $S=\{T_1, T_2, \cdots, T_s\}$, where $|S|=s$.
We further assume that $T'_k$ is a display of $T_k$ in $N$. 
Then,  each edge $e'_i$ of $T_k$ is mapped to a path $P'_i$
of $T'_k$, where $1\leq i\leq 2|X|-2$. Since $N$ displays $T_i$, we derive  the following linear equation system from the display of $T_k$:
\begin{eqnarray}
 \sum_{1\leq j\leq m} a_{ij}w(e_j)=w(e'_i), \;\; i=1, 2, \cdots,  2|X|-2,
 \label{eqn1}
\end{eqnarray}
where 
$$ a_{ij}=\left\{ \begin{array}{cc}
    1 & e_{j}\in E(P'_i); \\
    0 &  e_{j}\not\in E(P'_i).
\end{array}
\right.
$$
Let the coefficient matrix of  Eqn.~(\ref{eqn1}) be
 $A_k=(a_{ij})$,
which is a $(2|X|-2)\times m$ matrix, and let:
$$ W_k= \left( \begin{array}{c}
w(e'_1) \\ w(e'_2)\\ \vdots \\ w\left(e'_{2|X|-2}\right) \end{array} \right).$$
Since $N$ displays every tree of $S$,  we then determine the edge weights of $N$ by solving the following linear equation system:
\begin{eqnarray}
\label{eqn2}
  \left( \begin{array}{c}
A_1 \\ A_2\\ \vdots \\ A_s \end{array} \right) \times  \left( \begin{array}{c}
x_1 \\ x_2\\ \vdots \\ x_m \end{array} \right) 
=  \left( \begin{array}{c}
W_1 \\ W_2\\ \vdots \\ W_s \end{array} \right)
\end{eqnarray}
Note that Eqn.~(\ref{eqn2}) is a linear equation system that contains $2s(|X|-1)$ equations and at most $5|X|-4$ variable, as each $T_i$ contains $2|X|-2$ edges and $N$ contains 
$3r+2|X|-1$, where $r$ is the number of reticulations, which is at most $|X|-1$. 
\\

\begin{figure}[t!]
\centering
\includegraphics[scale=0.45]{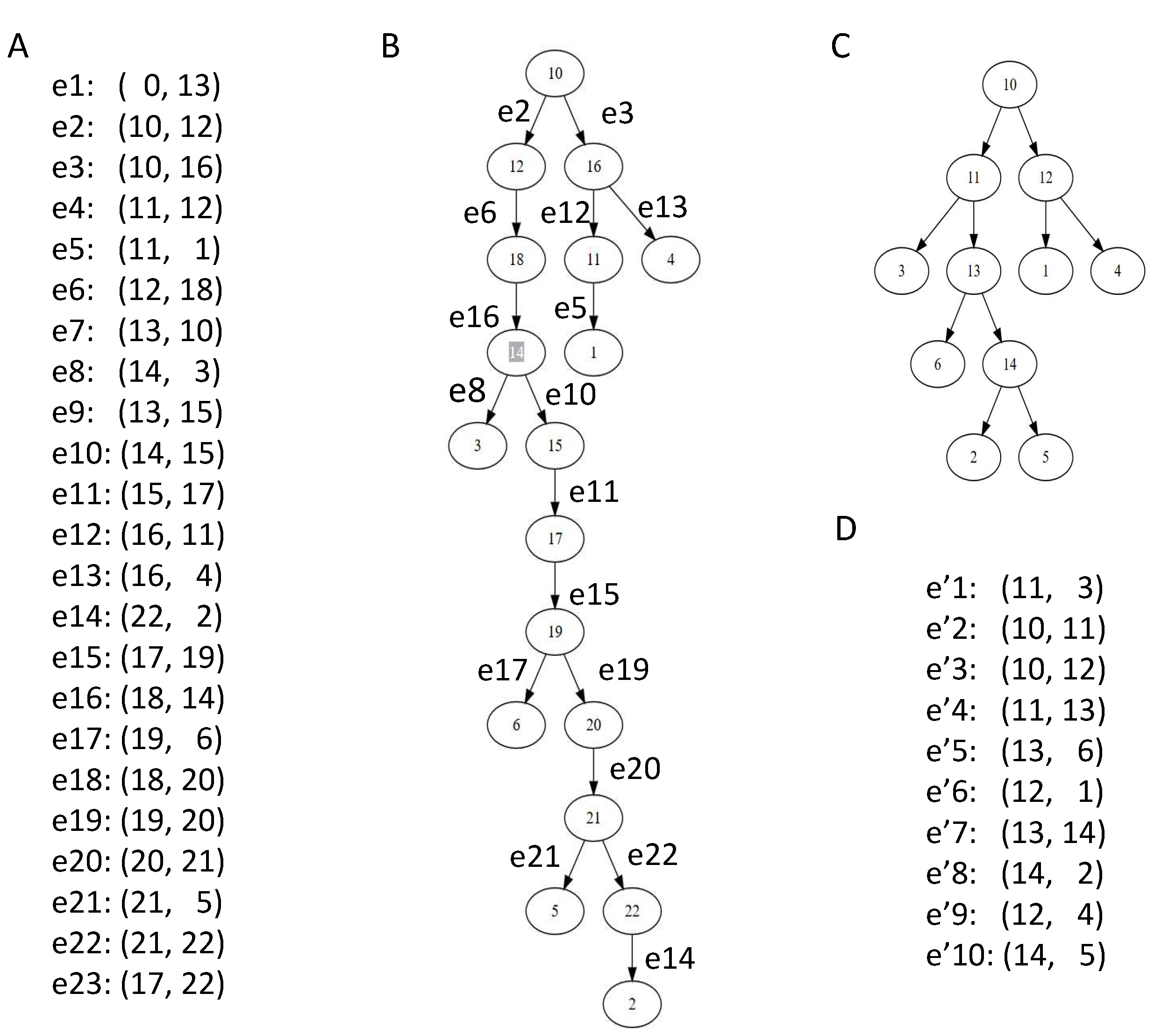}
\caption{An illustration of how to derive linear equations from a tree display.
(A) The list of the edges of a tree-child network. (B) A display of the tree in C.
(C) a phylogenetic tree on six taxa (1 to 6). (D) the list of the edges of the tree in $C$.
\label{FigA4}}
\end{figure}

\noindent {\bf Example 1.}  The edge list of a tree-child network is given 
in Figure~\ref{FigA4}A, where the full network is not given here. Figure~\ref{FigA4}B presents a particular display of the tree in Figure~\ref{FigA4}C, whose edges are listed in Figure~\ref{FigA4}D. 
In the display of the tree, the edge $e'_2$ is mapped to the path from the node 10 to the node 14, which consists of three edges 
$e2, e6, e16$ (Figure~\ref{FigA4}B). 
From $e'_2$ and its image, we obtain the following equation in the linear equation system Eqn.~(\ref{eqn2}):
 $$x_2+x_4+x_{16}=w(e'_2).$$

In general,  $N$ may not perfectly  display every $T$ when branch weights are considered. Therefore, let us set:
\begin{eqnarray}
  A= \left( \begin{array}{l}
A_1 \\ A_2\\\vdots \\ A_s \end{array}
 \right)\\
W=\left(\begin{array}{l}
W_1 \\ W_2\\\vdots \\ W_s \end{array} \right).
\end{eqnarray}
Noticing that  
$$\sum ^{s}_{i=1} ||T'_i-T_i||^2_2=||AX-W||^2_2,$$ 
we determine the branch weights of $N$ by solving the following quadratic optimization problem: 
\begin{eqnarray}
  &&\min ||AX-W||^2_2\\
  \textrm{subject to:} && \nonumber \\
&& x_j\geq 0,\;\; 1\leq j\leq m.
\end{eqnarray}

{\bf Remark.} Let $r$ be a reticulation node that has incoming $e_1, e_2, \cdots, e_d$ and the outgoing $e_{d+1}$.  For each input tree $T_i$, there is exactly one of edge pairs $(e_1, e_{d+1})$,  $(e_2, e_{d+1})$, $\cdots$,
$(e_d, e_{d+1})$ appearing in the display of $T_i$. Thus, solving the above optimization problem can only determine the value of  $w(e_i)+w(e_{d+1})$ for $i\leq d$.

\section*{C. A phylogenetic network for hominin relationships}
We analysed the morphological data in Dembo et al. 
(Proc Royal Soc B: Biol. Sci., vol. 282, 2015) by sampling 500 phylogenetic trees from a posterior collection of trees estimated from the morphological data. We computed the distance between each pair of trees using the rooted tree metric described in Kendall and Colijn (Mol. Biol. Evol.,  vol. 33, 2016). Briefly, this metric is the Euclidean distance between two vectors (one for each tree). The vector captures the amount of shared ancestry between each pair of tips, as well as each tip's distance from its parent. We used the tree topology only ($\lambda =0$ in the tree metric in the `treespace' function in the `treespace` package in R (Jombart et al., Mol. Ecol. Resour., vol. 17, 2017)). The amount of shared ancestry is the length of the path (in a phylogeny) between the root and the most recent common ancestor of a pair of tips.  Having found pairwise distances between all pairs of trees in our sample of 500, we clustered the trees into five clusters using Ward clustering. We chose two trees uniformly at random from each of the five clusters, as input for the analysis presented here.

\begin{figure*}[th!]
\centering
\includegraphics[scale=0.35]{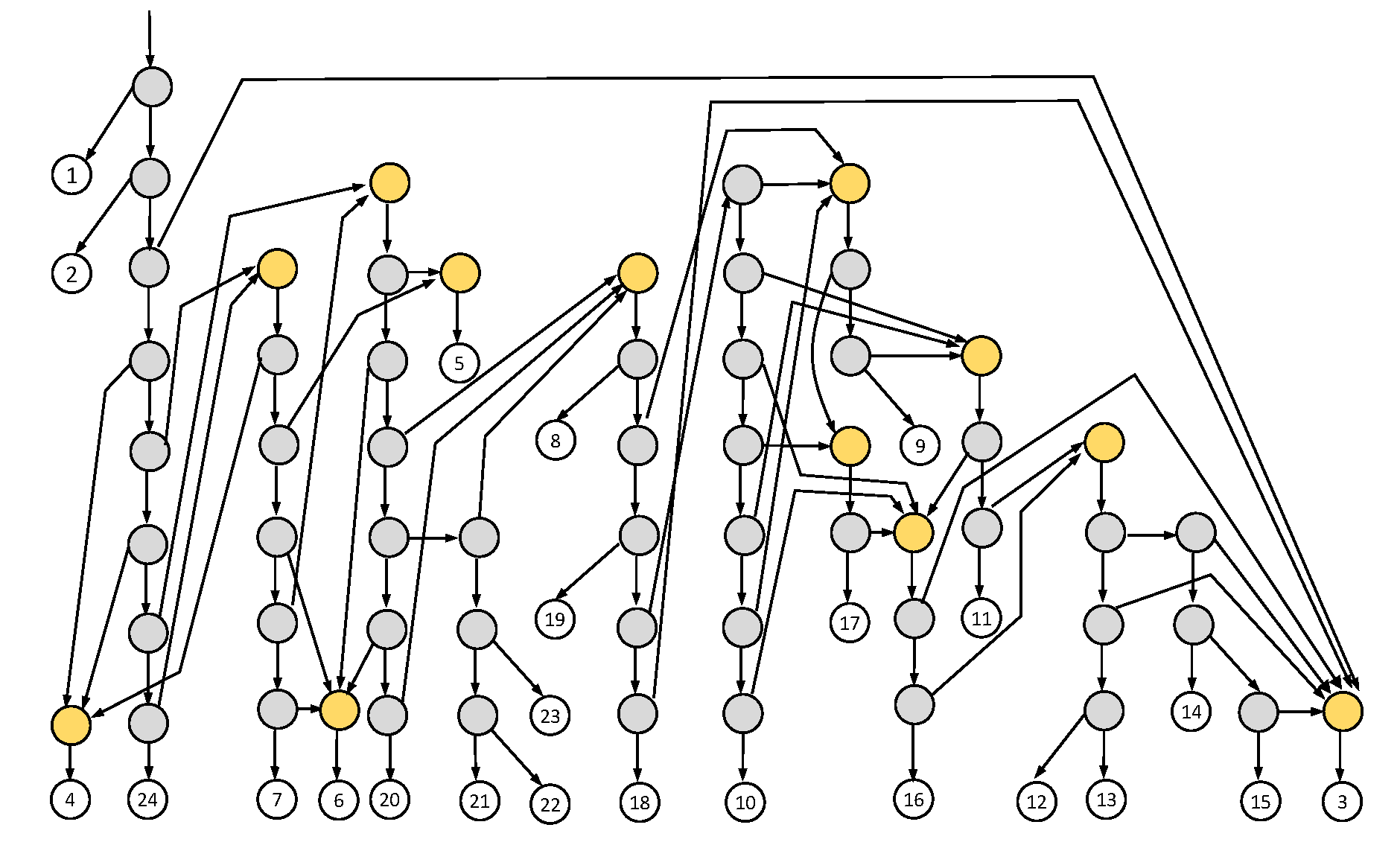}
\caption{\small A network model of hominin relationships.  1: {\it G.~gorilla}; 2: {\it P.~troglodytes};
3: {\it H.~floresiensis}; 4:  {\it   Ar.~ramidus};
5: {\it Au.~anamensis}; 6:  {\it Au.~afarensis};
 7: {\it K.~platyops}; 8: {\it Au.~africanus};
9: {\it Au.~sediba}; 10: African~{\it H.~erectus}; 
 11: Asian~{\it H.~erectus};
 12: {\it H.~heidelbergensis};
 13: {\it H.~neanderthalensis};
 14: {\it H.~sapiens};
 15: {\it H.~naledi};
 16: {\it H.~antecessor};
 17: Georgian~{\it H.~erectus};
 18: {\it H.~rudolfensis};
 19: {\it H.~habilis};
 20: {\it Au.~garhi};
 21: {\it P.~robustus};
 22: {\it P.~boisei};
 23: {\it P.~aethiopicus};
 24: {\it S.~tchadensis}.
\label{fig10_he}}
\end{figure*}

Hominins' phylogenetic relationships are not fully established. Due to the nature of the morphological data, the trees were discordant, and no single tree captures a highly-supported pattern of ancestry among the taxa. This motivates using a network to illustrate the complex ancestral relationships among these data.  Using  ALTS, we reconstructed a  network model  (Figure~\ref{fig10_he}) for hominin species using the 10 phylogenetic trees. 
%

The resulting network model contains 12 reticulation events. The top tree-node component contains the two outgroup species {\it  G. gorilla} and {\it P. troglodytes}, as well as the oldest hominin species, {\it S. tchadensis}. The three earliest members of the genus {\it Homo} ( African {\it H. erectus}, {\it H. rudolfensis} and {\it  H. habilis}),  together with  {\it  Au. africanus},  appear in a tree-node component, whereas four recent members of the genus {\it Homo} ({\it H. heidelbergensis},  {\it H. neanderthalensis},  {\it H. sapiens}
and {\it H. naledi}) compose another tree-node component. 
The three members of the genus 
{\it Paranthropus},  together with {\it Au. garhi},  compose a tree-node component.
The model also reflects the  high uncertainty about the phylogenetic position of {\it H. floresiensis}, who lived in the island of Flores, Indonesia (Argue et al., J Human Evol., vol. 57, 2009). 

\newpage
\thispagestyle{empty}
\centering
\includegraphics[scale=0.9]{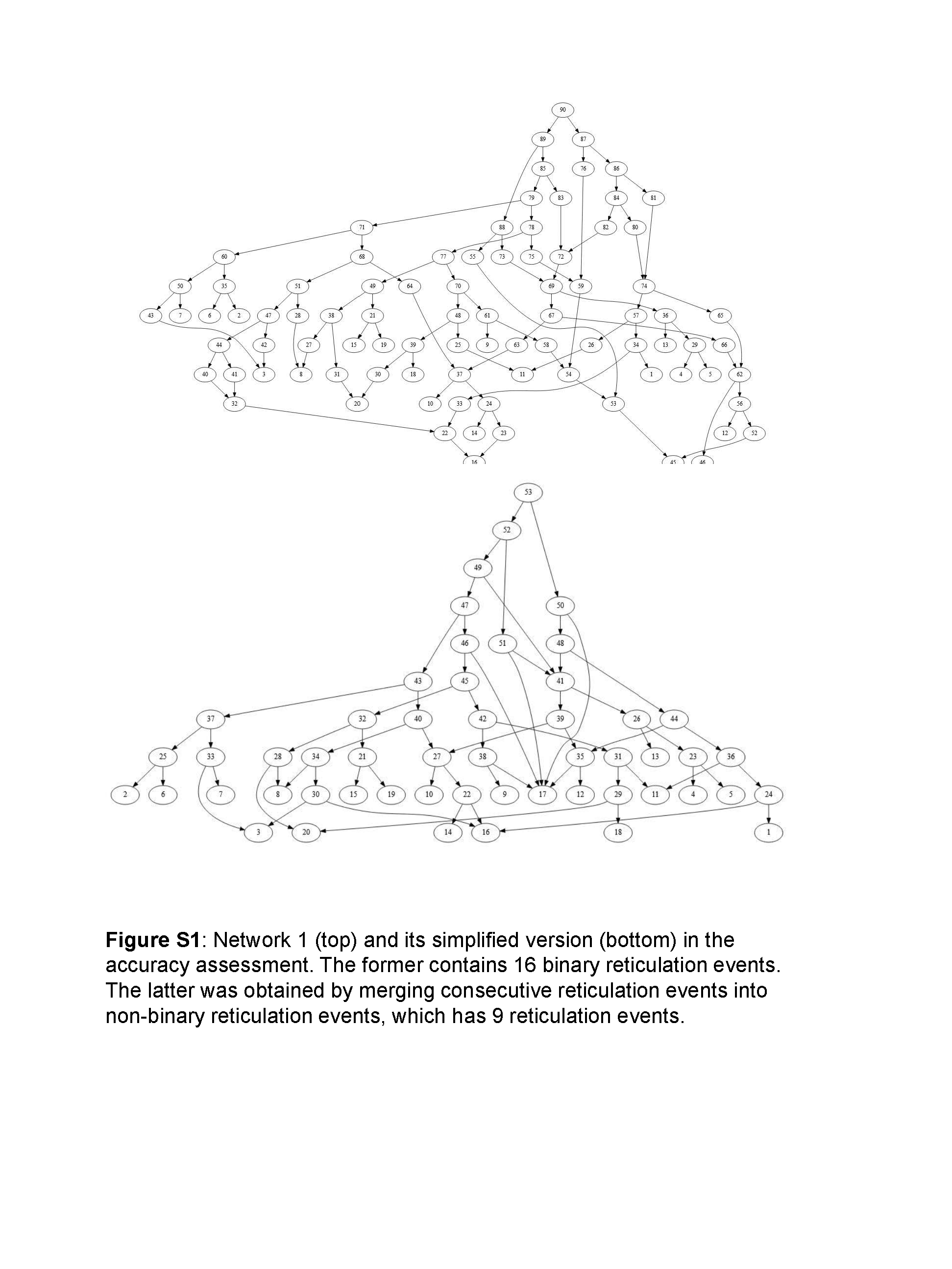}

\newpage 
\thispagestyle{empty}
\centering
\includegraphics[scale=0.8]{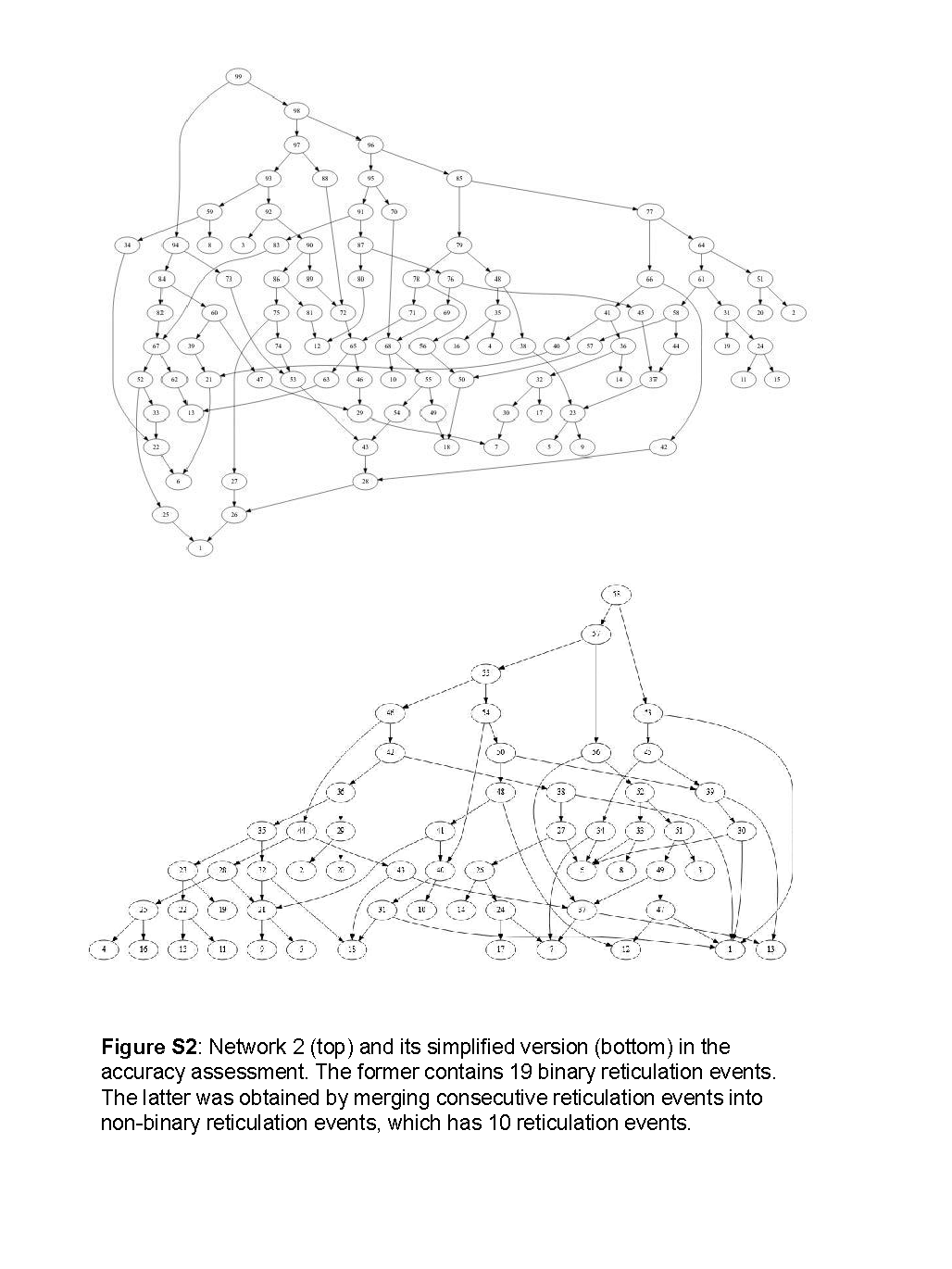}

\newpage 
\thispagestyle{empty}
\centering
\includegraphics[scale=0.9]{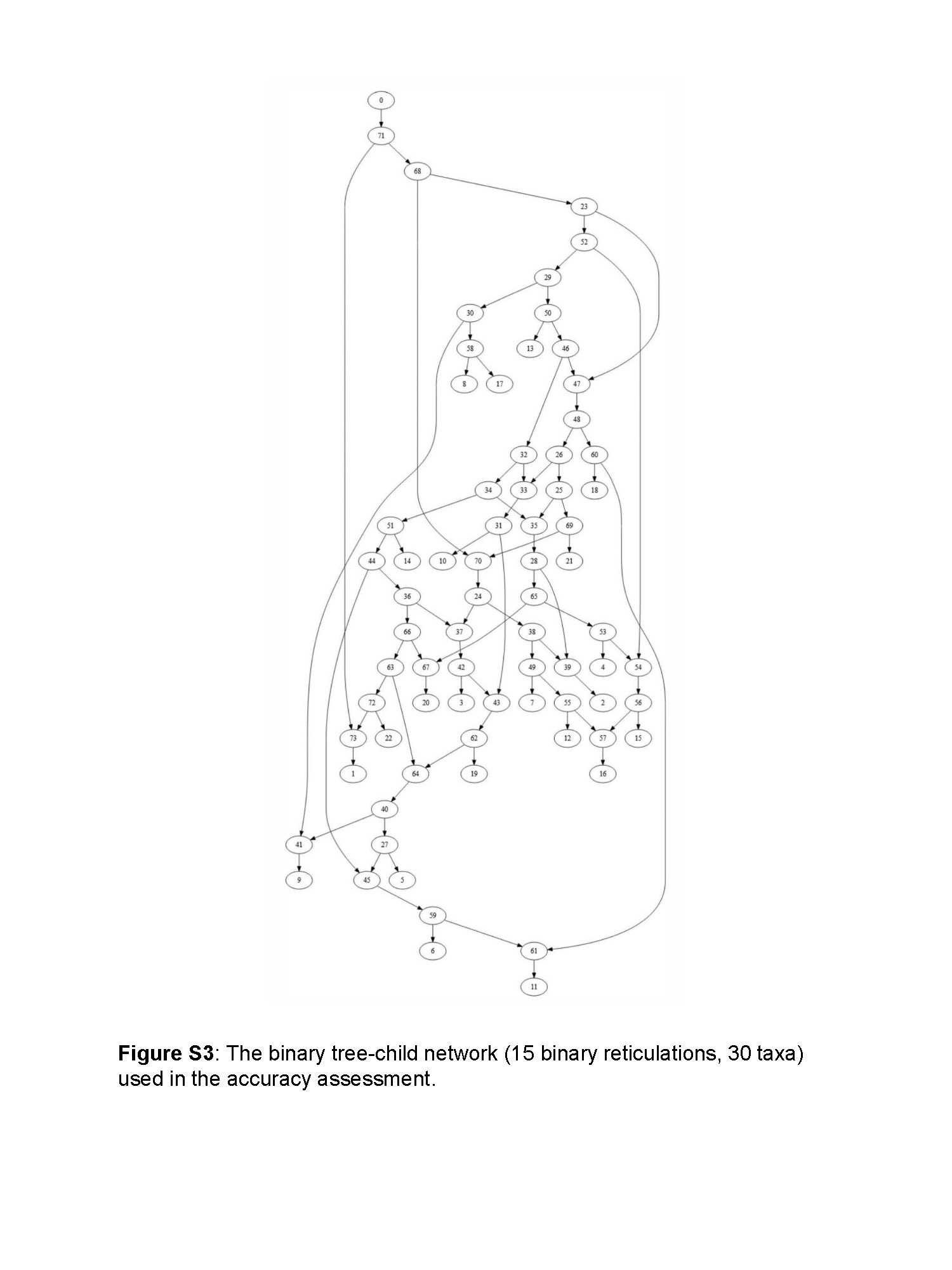}

\newpage
\thispagestyle{empty}
\centering
\includegraphics[scale=0.8]{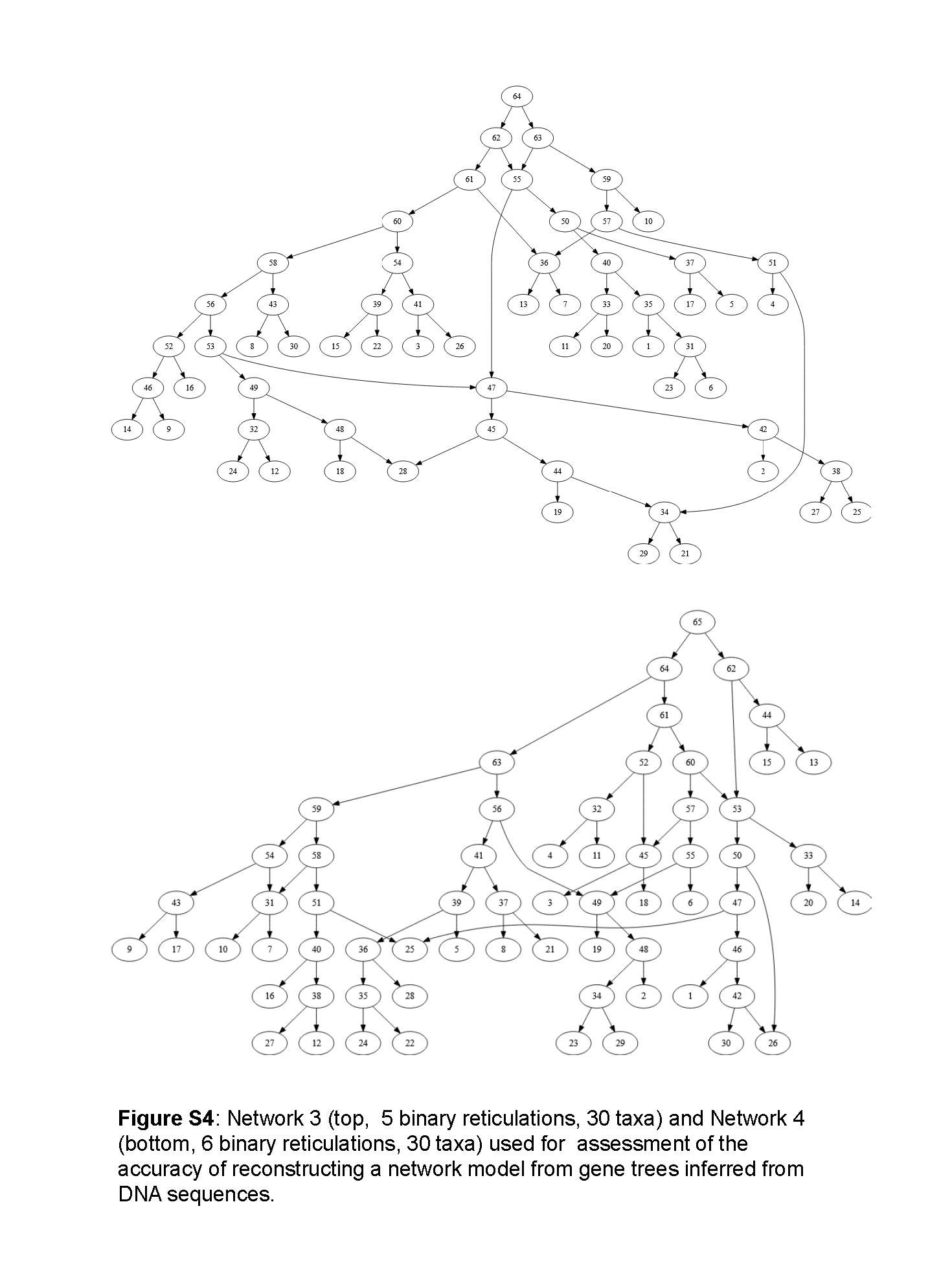}

\end{singlespace} 
\end{document}